\RequirePackage{ifpdf}
\documentclass[hyper]{JHEP3}

\pdfoutput=1
\usepackage{epsfig}
\usepackage{cite}
\usepackage{graphicx}
\usepackage{subfigure}
\usepackage{inputenc}
\usepackage{amsmath}

\newcommand{\beq}{\begin{equation}}
\newcommand{\eeq}{\end{equation}}
\newcommand{\beqn}{\begin{eqnarray}}
\newcommand{\eeqn}{\end{eqnarray}}
\def\df{{\rm d}}

\newcommand{\mcdot}{\!\cdot\!}
\newcommand{\lnQ}{{\rm Ln}_Q}
\newcommand{\lnQSq}{{\rm Ln}_Q^2}
\newcommand{\eq}[1]{Eq.~\eqref{eq:#1}}

\newcommand{\sect}[1]{Section~\ref{sec:#1}}
\newcommand{\apx}[1]{Appendix~\ref{app:#1}}
\def\a{\alpha}

\def\nn{\nonumber\\}
\def\pd{\partial}

\usepackage{amssymb}

\title{Combining initial-state resummation with fixed-order calculations of electroweak corrections}

\author{Christian W.~Bauer$^{a,b}$, Nicolas Ferland$^a$ and Bryan R.~Webber$^c$\\
   $^a$Ernest Orlando Lawrence Berkeley National Laboratory, University of California, Berkeley, CA 94720, USA\\
   $^b$Theoretical Physics Department, CERN, Geneva, Switzerland\\   
   $^c$University of Cambridge, Cavendish Laboratory, J.J.\ Thomson Avenue, Cambridge, UK\\
        E-mail: \email {cwbauer@lbl.gov}, \email{nferland@lbl.gov}, \email{webber@hep.phy.cam.ac.uk}
        }

\received{\today}               
\accepted{\today}               

\preprint{ Cavendish-HEP-17/14\\CERN-TH-2017-264}

\abstract{
We present a resummation of those double-logarithmically enhanced 
electroweak correction that arise in $pp$ colliders because protons 
are not SU(2) singlets, by solving DGLAP equations in the full
Standard Model.  We then show how to match these results with those of
fixed-order electroweak calculations.  At a 100 TeV $pp$ collider,
contributions beyond order $\alpha$ are $\sim 10$\% at partonic
center-of-mass energies of a few TeV.  These are mainly due to initial
states with massive vector bosons.
}

\begin{document} 

\section{Introduction}
\label{sec:introduction}

Throughout the history of particle physics, there has been a push to
probe fundamental interactions at shorter and shorter distance
scales. Many proposed future colliders would operate at energies
higher than those currently accessible: see, for
example~\cite{Arkani-Hamed:2015vfh,Mangano:2016jyj}. It is well known that
electroweak corrections grow double-logarithmically with the energy
scale of the partonic interaction, and a detailed understanding of
electroweak corrections is therefore important when trying to assess
the physics potential of future colliders and also to get precise
predictions at current colliders.

Double logarithmic corrections in exclusive processes, due to
suppression of gauge boson emission,  are familiar in both strong
and electroweak processes.  They can be resummed to all
orders, leading to Sudakov form factors. However, in electroweak
processes there are additional double logarithms that appear even in
observables and final states that are fully inclusive with respect to
extra boson emission~\cite{Ciafaloni:1998xg,Ciafaloni:1999ub,Ciafaloni:2000df,Ciafaloni:2000gm,Ciafaloni:2000rp,Ciafaloni:2001vu,Ciafaloni:2001mu,Ciafaloni:2003xf,Ciafaloni:2005fm,Ciafaloni:2006qu,Ciafaloni:2008cr,Ciafaloni:2009mm,Ciafaloni:2010ti,Forte:2015cia,Mangano:2016jyj,Bauer:2017isx}.
These are due to incomplete cancellation of logarithmic enhancements in
processes that are not symmetric with respect to weak isospin SU(2).
And since the beams of any current or proposed future collider are not
SU(2) symmetric, no observable measured at such colliders can be
symmetric, irrespective of how inclusively one defines the observable
and the final state.  For processes with non-symmetric
final states, such as charged lepton pair production, even if fully
inclusive with respect to electroweak boson emission, there will be
additional double logarithms.\footnote{The other extreme case where
  the observable is completely exclusive over the extra electroweak
  radiation has been studied many times
  before~\cite{Ciafaloni:1998xg,Ciafaloni:1999ub,Fadin:1999bq,Kuhn:1999nn,Denner:2000jv,Denner:2001gw,Feucht:2004rp,Jantzen:2005xi,Jantzen:2005az,Beccaria:2000jz,Hori:2000tm,Beenakker:2001kf,Denner:2003wi,Pozzorini:2004rm,Jantzen:2006jv,Chiu:2007yn,Chiu:2007dg,Chiu:2008vv,Chiu:2009yx,Chiu:2009yz,Chiu:2009mg,Chiu:2009ft,Fuhrer:2010eu}.}

Thus, in general every order of electroweak perturbation theory comes with two
extra powers of logarithms of the form $\ln Q^2 / m_V^2$, where $Q$
denotes the partonic energy scale of the process, while $m_V$
is a scale of order the masses of the $Z$ and $W$ bosons.
This means that electroweak perturbation
theory is always an expansion of the form
\begin{align}
\langle O \rangle = \langle O \rangle^{(0)} + \alpha_2 \lnQSq \langle
  O \rangle^{(1)} + \left[\alpha_2 \lnQSq \right]^2 \langle O
  \rangle^{(2)} + {\cal O}(\alpha_2^n \lnQ^{2n-m})
\,,
\end{align}
where
\begin{align}
\lnQ \equiv \ln\frac{Q^2}{m_V^2}
\end{align}
and $\alpha_2$ is the SU(2) coupling.
Electroweak (EW) perturbation theory, therefore, becomes badly convergent at  large
partonic energies. However, the convergence can be improved by
identifying the double-logarithmic terms and resumming them to all
orders.

In this paper, we present a way to resum double logarithms associated
with the asymmetry of the initial state, and to match the results with
those of fixed-order EW calculations.\footnote{For recent examples of NLO EW calculations
see~\cite{Kallweit:2014xda,Kallweit:2015dum,Kallweit:2017khh,Granata:2017iod,Chiesa:2017gqx}
and references therein.}  For this purpose,
we will study completely inclusive observables, which
are defined to sum over a completely SU(2) symmetric final state. The
example we use later in the paper is inclusive di-lepton production at
a $pp$ collider, which is defined to include a lepton-antilepton pair of a given
generation and any number of extra gauge bosons in the final state. So
to next-to-leading order (NLO) EW accuracy, this process sums over the final states $\ell^+
\ell^- (+V)$, $\ell^+ \nu_\ell (+V)$, $\bar \nu_\ell \ell^- (+V)$,
$\bar \nu_\ell \nu_\ell (+V)$, where $\ell$ denotes, for example, the
electron and $\nu_\ell$ the electron neutrino and the $(+V)$ denotes
the possible addition of a $\gamma$, $Z$ or $W^{\pm}$ boson. Since the
final state is SU(2) symmetric, the only SU(2) breaking effect is
coming from the fact that the initial state protons are not SU(2)
symmetric. The large logarithmic terms from the initial state
radiation can be resummed through a DGLAP
evolution~\cite{Gribov:1972ri,Dokshitzer:1977sg,Altarelli:1977zs} using the
interactions of the full Standard Model~\cite{Ciafaloni:2005fm}, which
was performed recently in \cite{Bauer:2017isx}. By performing the
DGLAP evolution to first order in electroweak effects, one sums all
double logarithms and a large class of the single logarithms, namely
those coming from the hard collinear parts of the splitting functions.
Not included are subleading logarithms such as those coming from
precise limits of integration and higher-order corrections to splitting functions
and running couplings. This resummation is called leading logarithmic (LL).~\footnote{A similar resummation of final-state logarithms in non-symmetric final
states could be performed though DGLAP evolution of electroweak
fragmentation functions but will not be implemented here.}

This DGLAP evolution uses SU(3) $\otimes$ U(1)$_{\rm em}$ for scales $q$
less than some matching scale $q_V$ of order $m_V$,
and the full unbroken SU(3) $\otimes$ SU(2) $\otimes$ U(1) for $q > q_V$. Performing this evolution 
up to the scale $Q$ of the process results in the PDFs
\begin{align}
f_A^{\rm SM}(x, Q)
\end{align}
for all SM parton species $A$.
Given these PDFs, the logarithms are resummed at leading logarithmic accuracy by writing
\begin{align}
\label{eq:sigmaLL}
\langle O \rangle_{\rm LL} = \sum_{AB}  \int \! \df \Phi_n \, O_n(\Phi_n) \, {\cal L}^{\rm SM}_{AB}(x_A, x_B; Q) B_{AB}(\widehat\Phi_n)
\,,
\end{align}
where $\widehat\Phi_n$ denotes the phase space of the partonic process,
$B_{AB}(\widehat\Phi_n)$ is the cross section for the process
initiated by partons $A$ and $B$ and $\df\Phi_n$ is the phase-space
element including their momentum fractions:
\begin{align}
\df\Phi_n = \df x_A\,\df x_B \,\df \widehat\Phi_n
\,.
\end{align}
$O(\Phi_n)$ denotes the value of the given observable calculated from
the phase space point $\Phi_n$, and
\begin{align}
\label{eq:LumiSM}
{\cal L}^{\rm SM}_{AB}(x_A, x_B; Q) = f_A^{\rm SM}(x_A, Q) \, f_B^{\rm SM}(x_B, Q) 
\end{align}
is the parton luminosity evaluated with the full SM PDFs. Note that
since the parton luminosity in the full SM has contributions from
initial states not usually present, such as electroweak gauge bosons, one
requires knowledge of partonic cross sections that are not usually
considered.

Which initial-state partons $A$ and $B$ are required depends on the 
partonic process (inclusive di-lepton production in our case) and how one counts powers of the coupling
constants. We summarize the scaling
of the various PDFs with the electroweak coupling in
Table~\ref{tab:PDFScaling}. Gluons obviously do not 
contribute at the order we are working. One can see that in the strict LL limit,
where one only requires to reproduce $\alpha^n \lnQ^{2n}$ terms, one
only needs to keep quarks in the initial state. However, transverse
vector bosons (the photon as well as massive vector bosons) are only
suppressed by one power of the logarithm, and their relative
contribution grows with increasing partonic center-of-mass energy. This
makes them phenomenologically quite relevant and we will keep them in
our analysis. Leptons, longitudinal gauge bosons\footnote{Note that
  longitudinal gauge bosons can become very important in situations
  where the partonic process is sensitive to non-gauge interactions,
  for example in Higgs and heavy quark production. In such cases one
  should include their effects in fixed order. An
  alternative approach is proposed in~\cite{Chen:2016wkt}. }
and Higgs bosons are further suppressed, and their contributions will
be neglected in the following discussion, although their effects,  together with
the Yukawa couplings to the top quark, have been kept in the solution to 
the evolution equations.
\begin{table}
\begin{center}
\begin{tabular}{|c|c|c|}
\hline
PDF & leading $\alpha$ power & log scaling\\
\hline
$q$ & 0 &  $\alpha^n \lnQ^{2n}$ \\
$g$ & 0 &  $\alpha^n \lnQ^n$ \\
$\gamma$ &  1 &  $\alpha^n \lnQ^{2n-1}$ \\
$V_T$ &  1 &  $\alpha^n \lnQ^{2n-1}$ \\
$V_L$ &    2 &  $\alpha^n \lnQ^{2n-2}$ \\
$\ell$ &  2 &  $\alpha^n \lnQ^{2n-2}$ \\
$h$ &   2 &  $\alpha^n \lnQ^{2n-2}$ \\
\hline
\end{tabular}
\end{center}
\caption{\label{tab:PDFScaling} 
The scaling of the PDFs with the EW coupling constant.}
\end{table}

Since the DGLAP evolution assumes the unbroken Standard Model (SM)
above the matching scale $q_V\sim m_V$, it drops all terms of order
$m_V / Q$,  which clearly misses important threshold effects around
the electroweak scale.\footnote{Some terms of this nature may be
  included by using modified splitting functions~\cite{Chen:2016wkt}.}
Furthermore, single logarithmic terms of order $\alpha \,
\lnQ$ are not fully accounted for in the DGLAP evolution. While these
effects do not need to be resummed for any scale $Q$ of interest,
at first order they can still give a relatively large effect and
introduce an uncertainty in the SM PDFs even for $Q \gg q_V$. One way
to estimate their importance is to vary the values of $q_V$ and $m_V$ chosen
in the DGLAP evolution, and it was shown in \cite{Bauer:2017isx} that
this can give an effect for certain PDFs at the 10\% level, even for
$Q \sim 10$  TeV. 

The threshold effects, as well as the single logarithmic terms, are properly included in any fixed-order EW calculation. This means that one way to obtain a result that includes the resummation of the LL logarithms, threshold effects, as well as single logarithmic terms is to combine a fixed-order EW calculation with the LL resummation. This is accomplished by the simple equation
\begin{align}
\label{eq:sigmaNLO_LL}
\langle O \rangle_{\rm NLO/LL} = \langle O \rangle_{\rm NLO} + \langle O \rangle_{\rm LL} - \left[ \langle O \rangle_{\rm LL}\right]_{\alpha}
\,.
\end{align}
Here $\langle O \rangle_{\rm NLO}$ denotes the fixed-order EW calculation at next-to-leading order, and $\left[ \langle O \rangle_{\rm LL}\right]_{\alpha}$ denotes the expansion of $\langle O \rangle_{\rm LL}$ in $\alpha$ up to the same order as included in the fixed-order expansion; in our case that requires an expansion to first order. This term is required to subtract the $O(1)$ and $O(\alpha)$ terms that are double counted between the NLO and the LL result.  It can be written as
\begin{align}
\label{eq:sigmaLL}
\left[ \langle O \rangle_{\rm LL}\right]_{\alpha} = \sum_{AB}  \int \! \df \Phi_n \, O_n(\Phi_n) \, \left[ {\cal L}^{\rm SM}_{AB}(x_A, x_B; Q)\right]_\alpha B_{AB}(\widehat\Phi_n)
\,,
\end{align}
where $\left[ {\cal L}^{\rm SM}_{AB}(x_A, x_B; Q)\right]_\alpha$ is the expansion of the SM parton luminosity. 

In summary, to combine a fixed-order EW calculation with the LL resummation of
the logarithms one requires only knowledge of the partonic cross sections $B_{AB}(\Phi_n)$ with $A,B$ including any SM particle (which are already required for the LL resummed result), as well as the expansion of the SM parton luminosity. We perform the calculation of the latter in \sect{partonLuminosities}, where we also study the convergence of the PDFs and parton luminosities in detail. In \sect{diLeptonProduction} we show the numerical impact of adding the LL resummation to a fixed order computation for the example of di-lepton production. We present our conclusions in \sect{conclusions}, and give the results of the required partonic cross sections in \apx{Partonic_Born}. 

\section{Standard Model parton luminosities and their expansion}
\label{sec:partonLuminosities}

The parton luminosities in the SM, as defined in \eq{LumiSM}, require
PDFs using the full SM  evolution. The corresponding DGLAP equations
are, in the notation of \cite{Bauer:2017isx},\footnote{However,
  contrary to \cite{Bauer:2017isx}, the PDFs here represent the actual
  momentum fraction distributions rather than the $x$-weighted distributions.}
 to leading order in all coupling constants
\beqn
\label{eq:genevol}
q\frac{\pd}{\pd q} f^{\rm SM}_i(x, q) &=& \sum_I
\frac{\alpha_{I}(q)}{\pi} \left[  P^V_{i,I}(q) \, f^{\rm SM}_i(x, q) +
  \sum_j  C_{ij,I} \int_x^{z_{\rm max}^{ij,I}(q)} \frac{\df z}z \,
  P^R_{ij, I}(z) f^{\rm SM}_j(x/z, q) \right],\nn
\eeqn
where the sum over $I$ goes over all possible different
interactions\footnote{In this paper we neglect Yukawa interactions and
  the Higgs self-interaction, which make only very small
  contributions.} in the
Standard Model: $I=1$ for U(1), $I=2$ for SU(2), $I=3$ for SU(3)
and $I = M$ for mixed interactions proportional to
$\alpha_M(q) = \sqrt{\alpha_1(q)\, \alpha_2(q)}$.  The latter represent
interference between processes initiated by U(1) and SU(2) bosons. 
As in \cite{Bauer:2017isx} we choose $q_V = m_V = 100$ GeV. 
Note that one might want to go to higher orders in the QCD evolution,
and for that one can use the known higher-order splitting kernels. 

Since the evolution in the unbroken Standard Model only applies for scales $q > q_V$, one requires a boundary  condition at the scale $q_V$, which we write as
\begin{align}
f^{\rm SM}_i(x, q_V) = f^{\rm noEW}_i(x, q_V)
\,.\end{align}
The precise definition of $f^{\rm noEW}$ required depends on what
level of accuracy is desired. 
One clearly needs to include the QCD evolution from the hadronic scale
$q_0\sim 1$ GeV to the scale $q_V$, since $\alpha_s \ln q_V / q_0 \sim 1$. QED evolution gives rise to single logarithmic effects, and by including this evolution one includes terms of order $\alpha^n \ln^n q_V / q_0$, which should be subdominant to the double logarithmic terms generated by the EW evolution above $q_V$. However, by including this evolution also below $q_V$ one is using ${\cal O}(\alpha)$ evolution both above and below $q_V$. For this reason, we choose as boundary condition
\begin{align}
f^{\rm noEW}_i(x, q_V) = f^{\rm QCED}_i(x, q_V)
\,,\end{align}
where the PDF set QCED is obtained by $SU(3) \otimes U(1)_{\rm em}$
evolution from scales below $q_V$.  Specifically, as in
\cite{Bauer:2017isx}, we use the CT14qed PDF
set~\cite{Schmidt:2015zda} at 10 GeV and replace the
photon PDF by that of the LUXqed
set~\cite{Manohar:2016nzj,Manohar:2017eqh}.
The strongly interacting partons are rescaled to obtain exact momentum
conservation. The resulting PDF set is then evolved up to the matching scale
$q_V=100$ GeV using leading-order (LO) DGLAP equations that include
QCD and QED effects.
In this way we obtain a LO PDF set at the matching scale which is
consistent with our LO evolution above that scale.

The first contribution in \eq{genevol}, proportional to $P^V_{i,I}$, denotes the virtual contribution to the PDF evolution (the disappearance
of a flavor $i$), while the second contribution is the real contribution (the appearance of flavor $i$ due to the 
splitting of a flavor $j$). The maximum value of $z$ in the integration of the real contribution depends on the 
type of splitting and interaction
\beq
\label{eq:zmax}
z_{\rm max}^{ij,I}(q) = \left\{
\begin{array}{ll}
1 - \frac{m_V}{q} & {\rm for}\, I = 1, 2, \,{\rm and}\, i, j \notin V \,{\rm or}\, i, j \in V
\\
1 & {\rm otherwise}
\end{array}
\right.
\,,\eeq
This implies that an infrared cutoff $m_V$ is applied when a U(1)
boson $B$ or SU(2) boson $W$ is 
emitted. The physical origin of this cutoff is that the energy of a massive vector 
boson is bounded by its mass\footnote{Note that the precise value of
  the mass $m_V$ 
does not matter at LL accuracy. In \cite{Bauer:2017isx} the effect of
varying $m_V$ by a factor of 2 was used to obtain an estimate
of the uncertainties from higher logarithmic effects.}. It is also
required, since contrary to the standard SU(3)
and U(1) evolution equations, which are regular as $z \to 1$ due to a
cancellation between real and virtual contributions, the SU(2)
evolution equations are not regular as $z \to 1$ due to the
non-singlet nature of the initial state.

Before we expand the resulting PDFs, it is worth recalling where the
double-logarithmic sensitivity is coming from, since this is not
present in the usual DGLAP evolution.
One can understand this by looking for example at the evolution of an
up-type left-handed fermion due to the SU(2) interaction:
\beq
\label{eq:SU2_fermion}
q\frac{\pd}{\pd q}f^{\rm SM}_{u_L} =
\frac{\a_2}{\pi}
\int_0^{1-\frac{m_V}{q}} \frac{\df z}{z} P^R_{ff,G}(z) \left[ \frac{f^{\rm SM}_{d_L}(x/z, q)}{2}+\frac{f^{\rm SM}_{u_L}(x/z, q)}{4} -  \frac{3zf^{\rm SM}_{u_L}(x, q)}{4} \right]+\ldots
\eeq
where the terms $\ldots$ do not contribute to double logarithms.
The splitting function $P^R_{ff,G}(z)$ is singular as $z \to 1$. If the initial
state were SU(2) symmetric, one would have $f_{u_L}(x, q) = f_{d_L}(x,
q) \equiv f_{Q_L}(x, q)$ and the combination in the square bracket
would be of the form $3/4 \left[ f_{Q_L}(x/z, q) -  z\,f_{Q_L}(x,
  q)\right]$, such that the divergence in $z\to1$ would cancel in the
difference. Since $f_{u_L}(x, q) \neq f_{d_L}(x, q)$, this
cancellation does not happen, generating logarithmic sensitivity to
the ratio $m_V / q$ from the integral over $z$. This soft dependence
gives rise to the double logarithmic sensitivity in the solution of
the DGLAP equation. As was shown in \cite{Ciafaloni:2005fm,Bauer:2017isx}, in a basis
of definite weak isospin, this double logarithmic sensitivity drives
any terms with non-zero isospin to zero as $q \to \infty$, thereby
restoring EW symmetry asymptotically. For the PDFs
included, we will retain all DGLAP effects, even those that do not
give rise to double-logarithmic terms.  

As explained earlier, our aim is to obtain not only the luminosities
resulting from the resummed SM PDFs but also their expansion to  first
order in $\alpha_I$. This will permit matching to exact fixed-order
calculations and assessment of the contribution of terms
beyond fixed order.   To expand the PDFs to first order in $\alpha_{I\ne 3}$ we define
\begin{align}
\label{eq:gdef}
\left[f_i^{\rm SM}(x,q)\right]_\alpha = f_i^{\rm noEW}(x,q) + g_i(x,q)
\end{align}
such that $\left[f_i^{\rm SM}(x,q)\right]_\alpha$ only includes the linear terms in $\alpha_{I\ne 3}$. This implies
\begin{align}
f_i^{\rm SM}(x,q) = \left[f_i^{\rm SM}(x,q)\right]_\alpha + {\cal O}(\alpha_{I\ne 3}^2)\,.
\end{align}
The boundary condition for $g_i$ is 
\begin{align}
\label{eq:gBoundary}
g_i(x, q < q_V) = 0
\,.\end{align}
The definition of the function $g_i(x,q)$ obviously depends on the
definition of $f_i^{\rm noEW}(x,q)$.  The function $g_i$ vanishes for
$q < q_V$,  so $f_i^{\rm noEW}$ coincides with $f_i^{\rm SM}$ for
those values. Since the SM evolution for $q > q_V$ is adding the full
SU(2) $\otimes$ U(1) evolution, it makes sense to choose $f_i^{\rm
  noEW}(x,q)$ to  only include the SU(3) evolution above that
scale. In other words, we choose
\begin{align}
\label{eq:fNoEWDef}
f_i^{\rm noEW}(x,q) = \left\{ \begin{array}{ll} {\rm QCED \, evolution} & q < q_V\,, \\ {\rm QCD \, evolution} & q > q_V\,. \end{array} \right.
\end{align}
One could also choose a definition that includes the QED evolution for $q > q_V$. This would introduce spurious single logarithmic $[\alpha \lnQ]^n$ terms in the difference between $f_i^{\rm SM}$ and $[f_i^{\rm SM}(x,q)]_\alpha$, which are in principle beyond the claimed accuracy. However,  the definition \eq{fNoEWDef} trivially avoids these spurious terms, which is why it is our choice for the remainder of this paper.

The DGLAP equation for $[f_i^{\rm SM}(x,q)]_\alpha$ can easily be obtained by expanding \eq{genevol} to obtain
\begin{align}
\label{eq:genevolExp}
&q\frac{\pd}{\pd q} \left[f_i^{\rm SM}(x,q)\right]_\alpha
\nn
& \qquad
= \frac{\alpha_{3}(q)}{\pi} \left[  P^V_{i,3}(q) \, \left[f_i^{\rm SM}(x,q)\right]_\alpha +  \sum_j  C_{ij,I} 
\int_x^{1} \frac{\df z}z  P^R_{ij, 3}(z) \left[f_i^{\rm SM}(x/z,q)\right]_\alpha \right] 
\\
& \qquad +  \sum_{I\in 1,2,M}  \frac{\alpha_{I}(q)}{\pi} \left[  P^V_{i,I}(q) \, f^{\rm noEW}_i(x, q) +  \sum_j  C_{ij,I} 
\int_x^{z_{\rm max}^{ij,I}(q)} \frac{\df z}z  P^R_{ij, I}(z) f^{\rm noEW}_j(x/z, q) \right]\,.
\nonumber
\end{align}
In other words, we have simply set $[f_i^{\rm SM}(x,q)]_\alpha = f_i^{\rm noEW}$ in the second line, since the dropped terms give rise to second order effects. 
This gives
\begin{align}
\label{eq:genevolExp2}
& q\frac{\pd}{\pd q} g_i(x, q) 
\nn
&\qquad = \frac{\alpha_{3}(q)}{\pi} \left[  P^V_{i,3}(q) \, g_i(x, q) +  \sum_j  C_{ij,I} 
\int_x^{1} \frac{\df z}z \, P^R_{ij, 3}(z) g_j(x/z, q) \right] 
\\
&\qquad +  \sum_{I\in 1,2,M}  \frac{\alpha_{I}(q)}{\pi} \left[  P^V_{i,I}(q) \, f^{\rm noEW}_i(x, q) +  \sum_j  C_{ij,I} 
\int_x^{z_{\rm max}^{ij,I}(q)} \frac{\df z}z \, P^R_{ij, I}(z) f^{\rm noEW}_j(x/z, q) \right]\,.
\nonumber
\end{align}

We have implemented the DGLAP equation~\eq{genevolExp2} with boundary
condition~\eq{gBoundary} and solved for $g_i(x, q)$. 
As a cross check on the resulting expanded PDFs one can validate that the result is indeed linear in the coupling constants $\alpha_{I=1,2,M}$. For this, we perform the rescaling $\alpha_I \to r \alpha_I$, and then plot $[f_i^{\rm SM}(x,q)]_\alpha$ for various values of $r$ (normalized to the result with $r=1$). Figure~\ref{fig:scaling} clearly verifies the expected linear behavior.
\FIGURE[h]{
 \centering
  \includegraphics[scale=0.29]{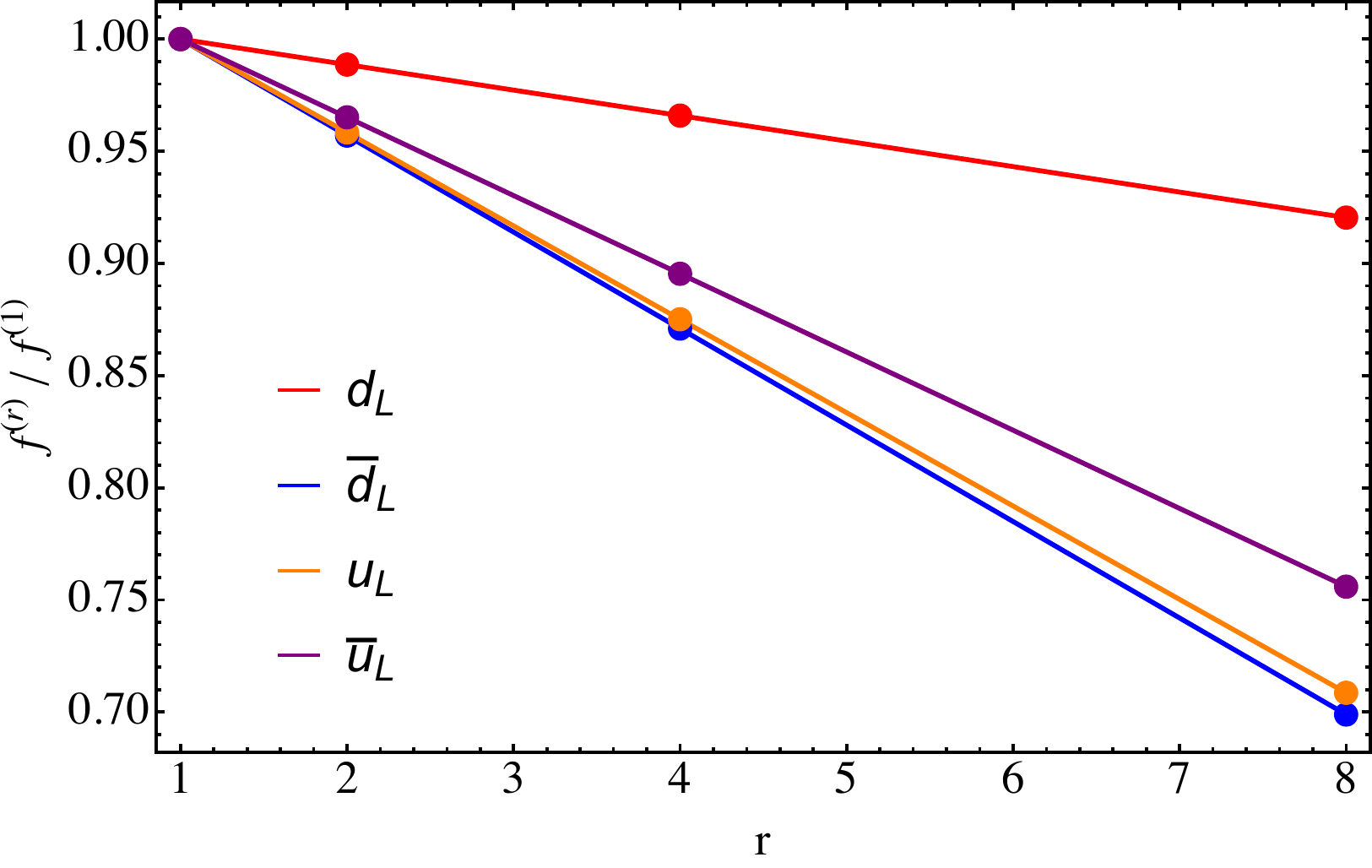}
  \includegraphics[scale=0.29]{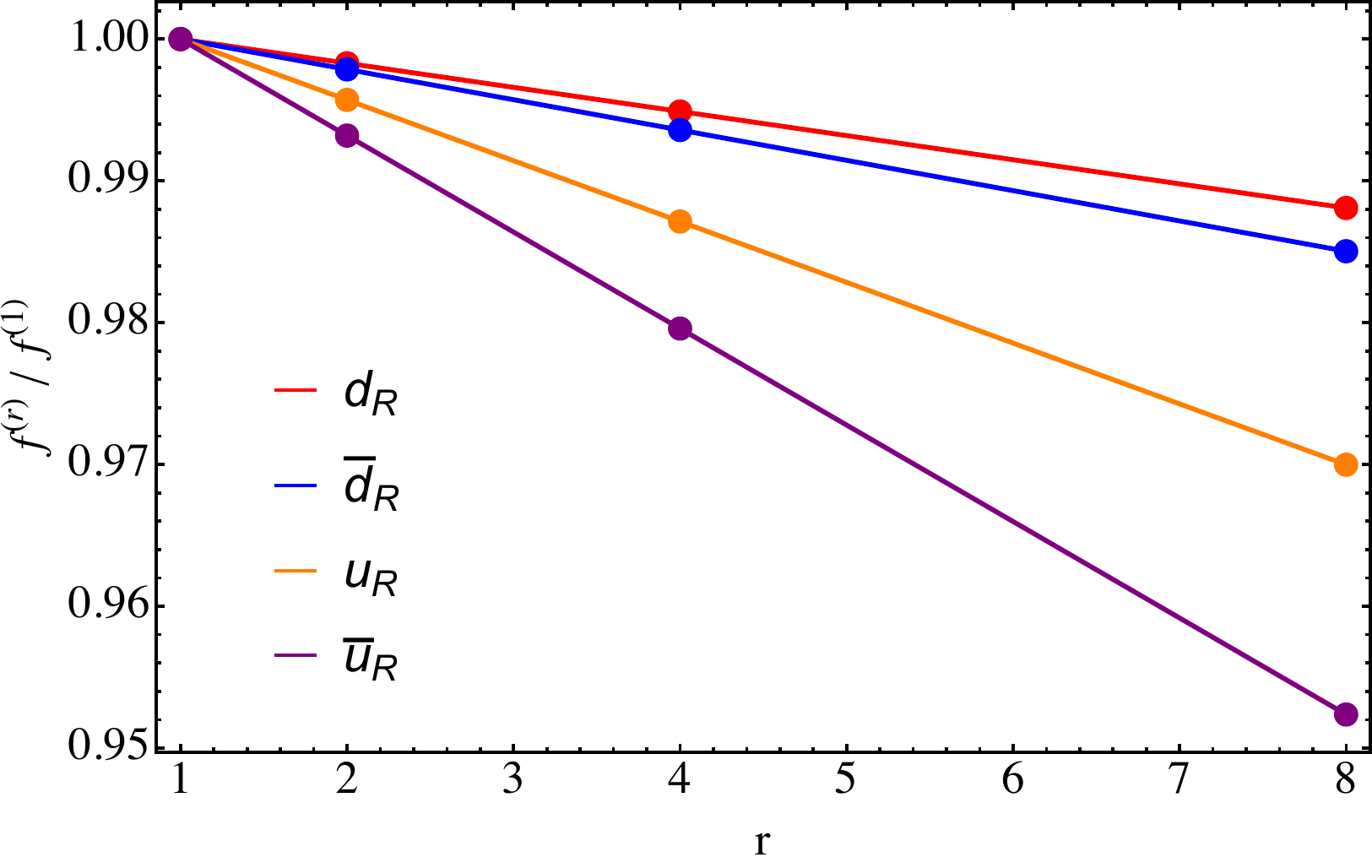}
  \includegraphics[scale=0.29]{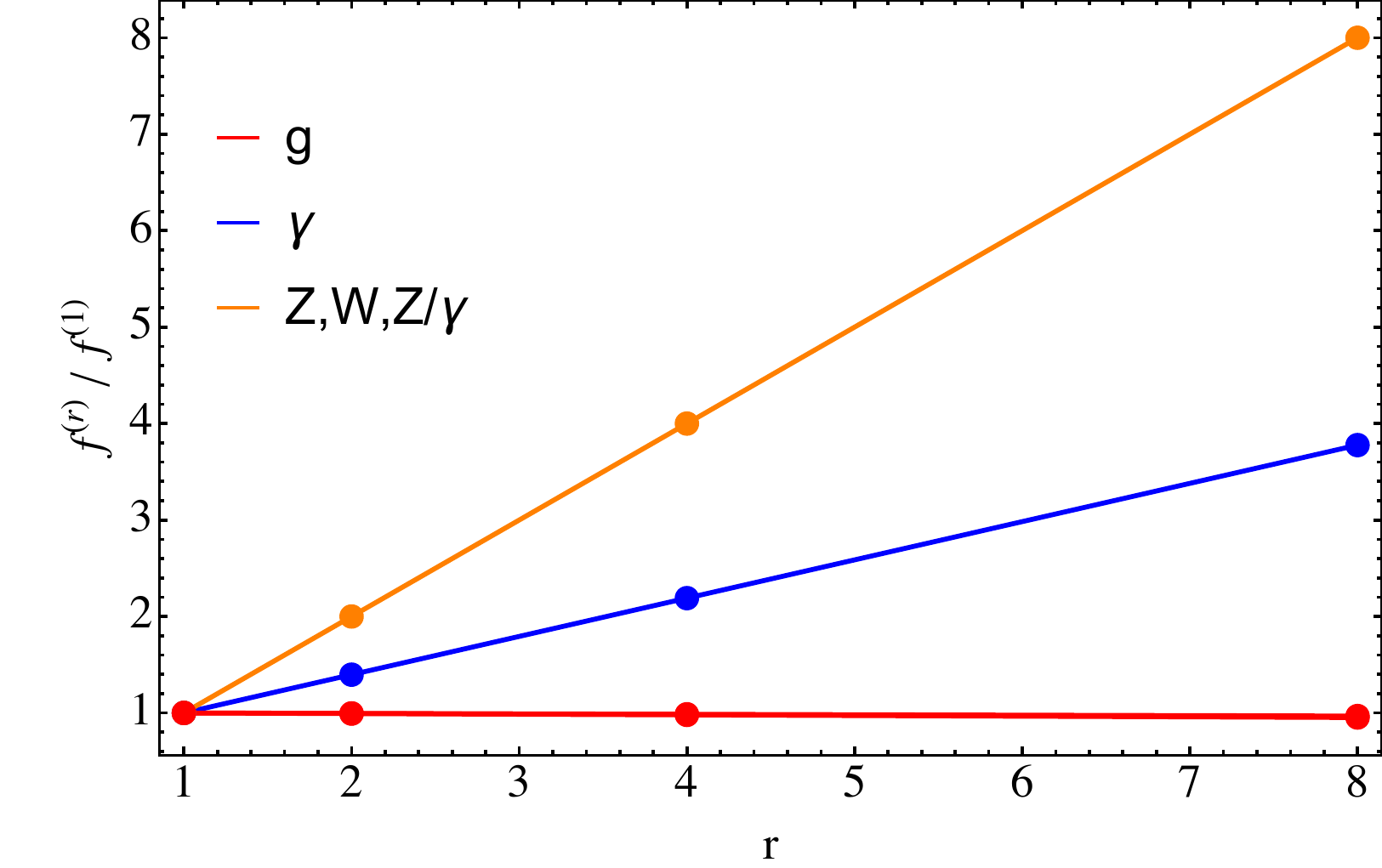}
\caption{\label{fig:scaling}%
Scaling of the expanded PDFs with the parameter $r$, which multiplies $\alpha_{I=1,2,M}$. On the left, we show the left-handed quarks of the first generation, in the middle the right-handed quarks of the first generation, and on the right the vector bosons. One can clearly see that the expanded PDFs are linear in $\alpha_I$. 
}}

Given the resummed result for the SM PDFs, together with this first-order expansion, one can obtain a first estimate of the higher-order effects, and the convergence of electroweak perturbation theory. For this, we define the two ratios
\begin{align}
r_i^{\rm noEW}(x,q) \equiv \frac{f^{\rm noEW}_i(x, q)}{f^{\rm SM}_i(x, q)}\,,
\qquad 
r_i^{\rm SM, \alpha}(x,q) \equiv \frac{\left[f_i^{\rm SM}(x,q)\right]_\alpha}{f^{\rm SM}_i(x, q)}
\,.\end{align}
Defining the function $h_i(x,q)$ to be the difference between $[f_i^{\rm SM}]_\alpha$ and $f_i^{\rm SM}$ we can write
\begin{align}
f_i^{\rm SM}(x,q) = f_i^{\rm noEW}(x,q) + g_i(x,q) + h_i(x,q)
\,,
\end{align}
where $g_i(x,q)$ is the same function used in \eq{gdef}. As already
discussed, the function $g_i(x,q)$ is of order $\alpha_I$, while the
function $h_i(x,q)$ contains the resummed terms of $\alpha_I^2$ and higher. 
With these definitions, one can write
\begin{align}
r_i^{\rm noEW}(x,q) &= 1 - \frac{g_i(x,q) + h_i(x,q)}{f^{\rm SM}_i(x, q)} \sim 1 + {\cal O}(\alpha_I)\,,
\nn
r_i^{\rm SM, \alpha}(x,q) &= 1 - \frac{h_i(x,q)}{f^{\rm SM}_i(x, q)} \sim 1 + {\cal O}(\alpha_I^2)
\,.
\end{align}
Thus, the deviation from unity of the first ratio shows the size of the first-order correction, while the deviation of the second ratio shows the size of the higher-order corrections. Note that for PDFs for which $f^{\rm noEW}_i(x, q)$ vanishes (in our case the massive vector bosons) the first ratio vanishes, and the second ratio gives
\begin{align}
\label{eq:ratio_VM}
r_i^{\rm SM, \alpha}(x,q) &= 1 - \frac{h_i(x,q)}{g_i(x,q)} \sim 1 + {\cal O}(\alpha_I)
\end{align}
and is therefore an estimate of the size of the second-order term relative to the first-order term. 

\FIGURE[h]{
 \centering
  \includegraphics[scale=0.4]{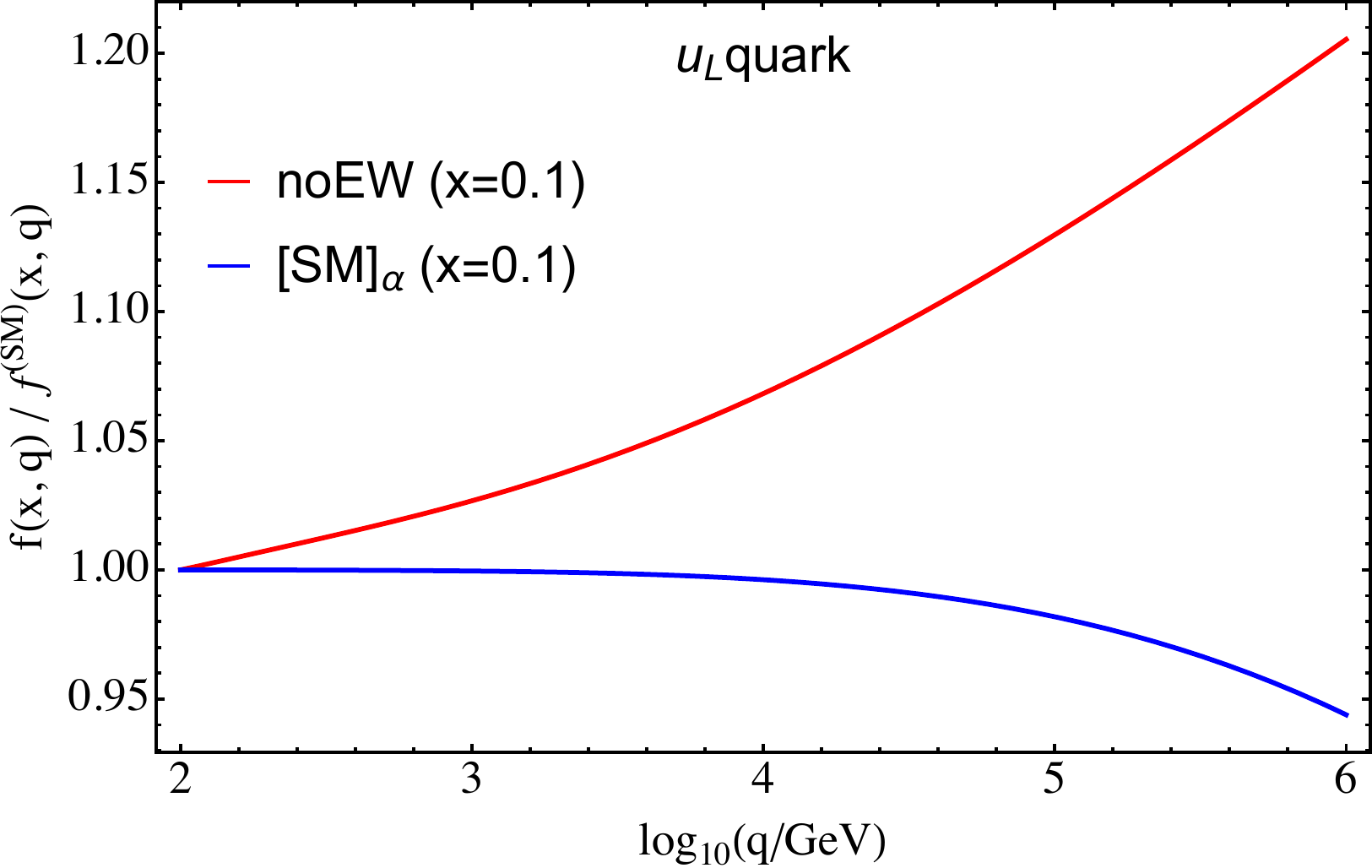}
  \includegraphics[scale=0.4]{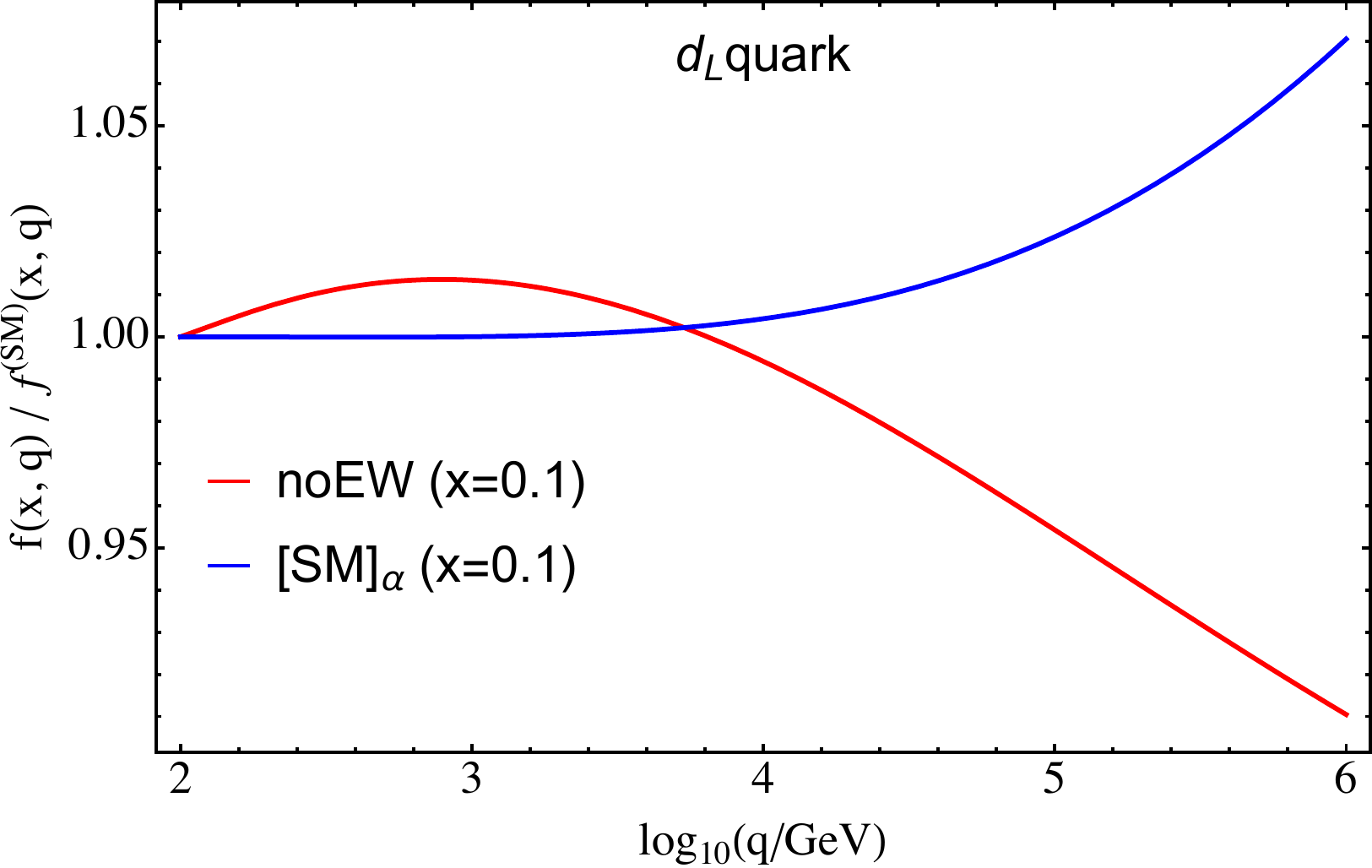}
  \includegraphics[scale=0.4]{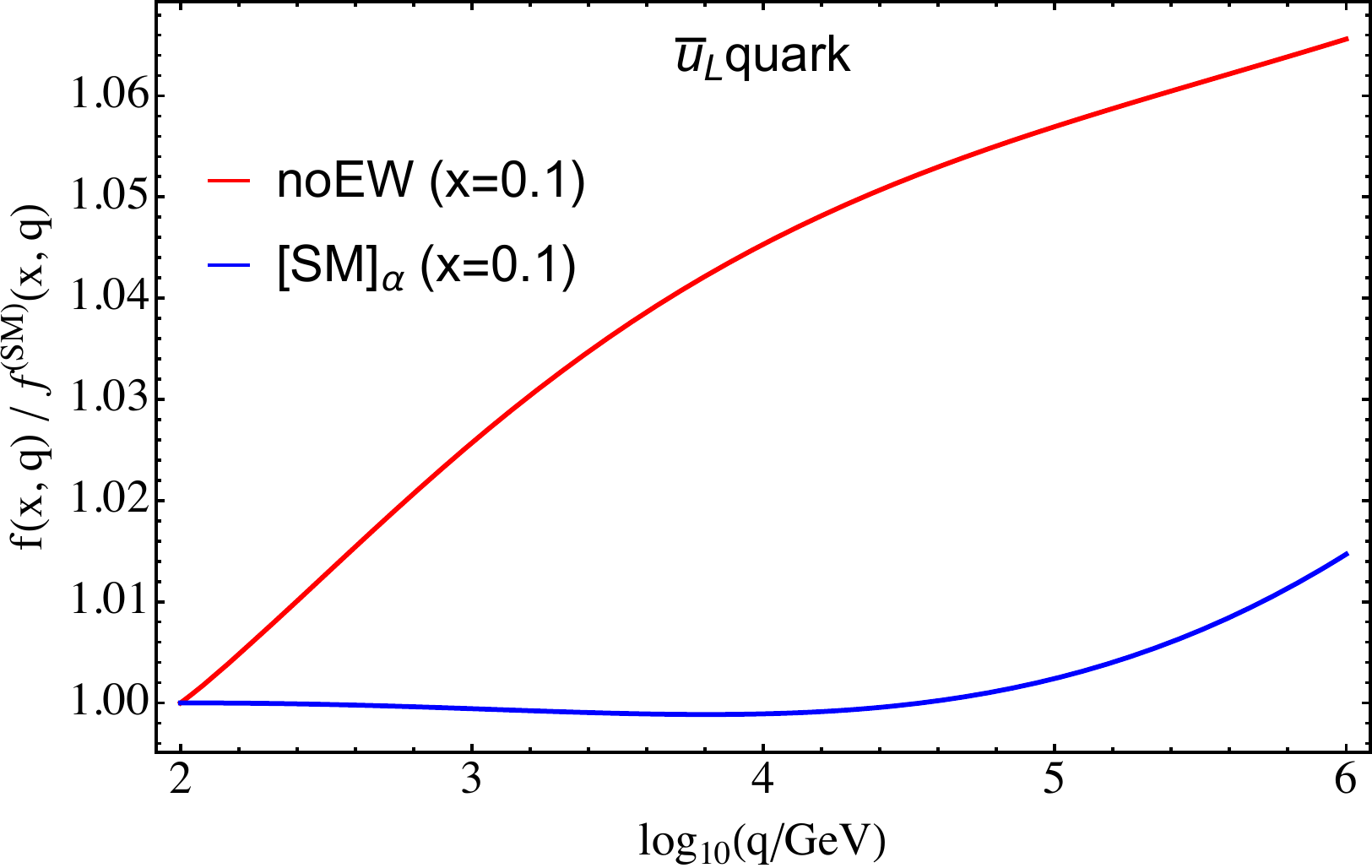}
  \includegraphics[scale=0.4]{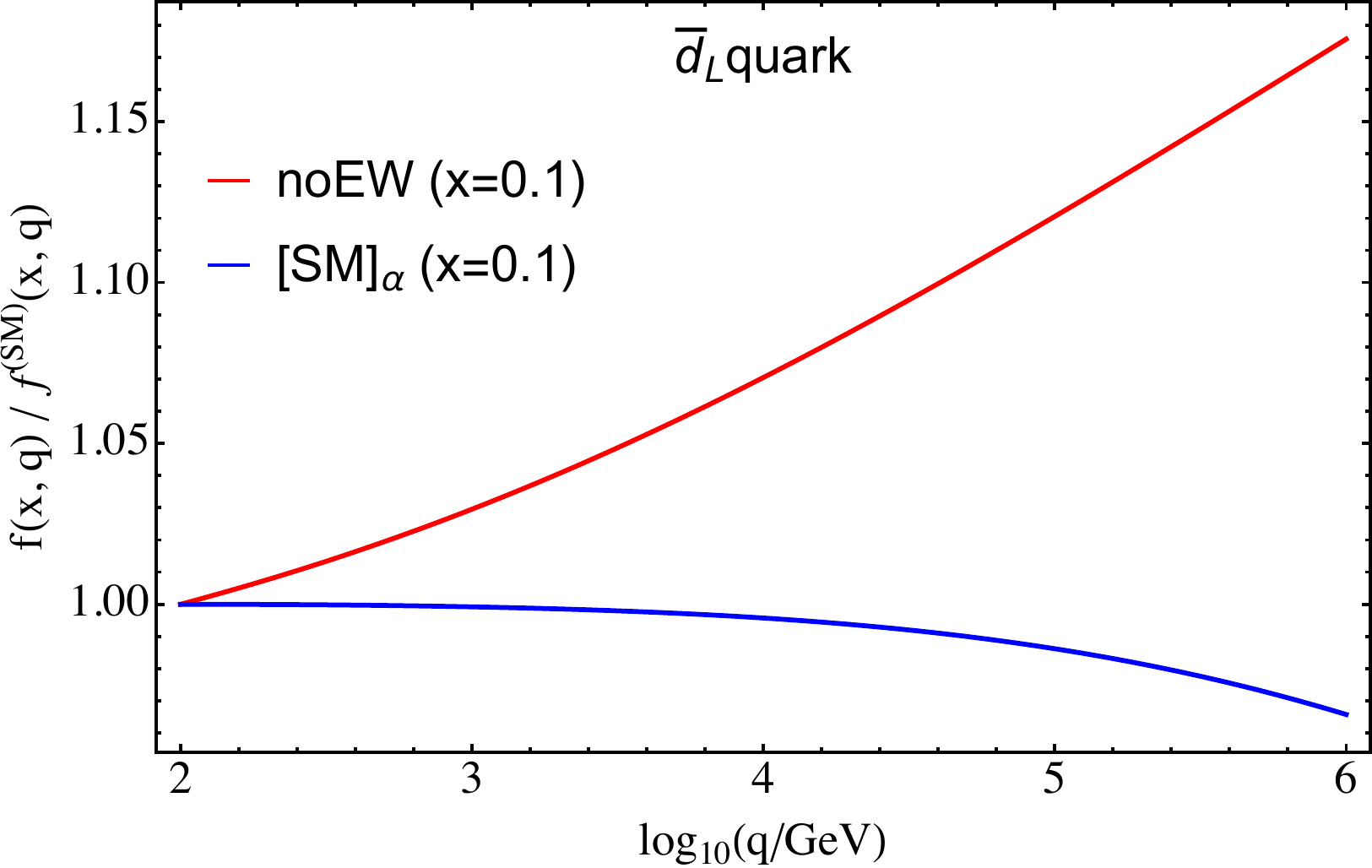}
\caption{\label{fig:PDFExpansion_Q}%
The ratio of the ``noEW'' and expanded SM PDFs relative to the PDF evaluated in the full SM for left-handed quarks.  
}}
\FIGURE[h]{
 \centering
  \includegraphics[scale=0.4]{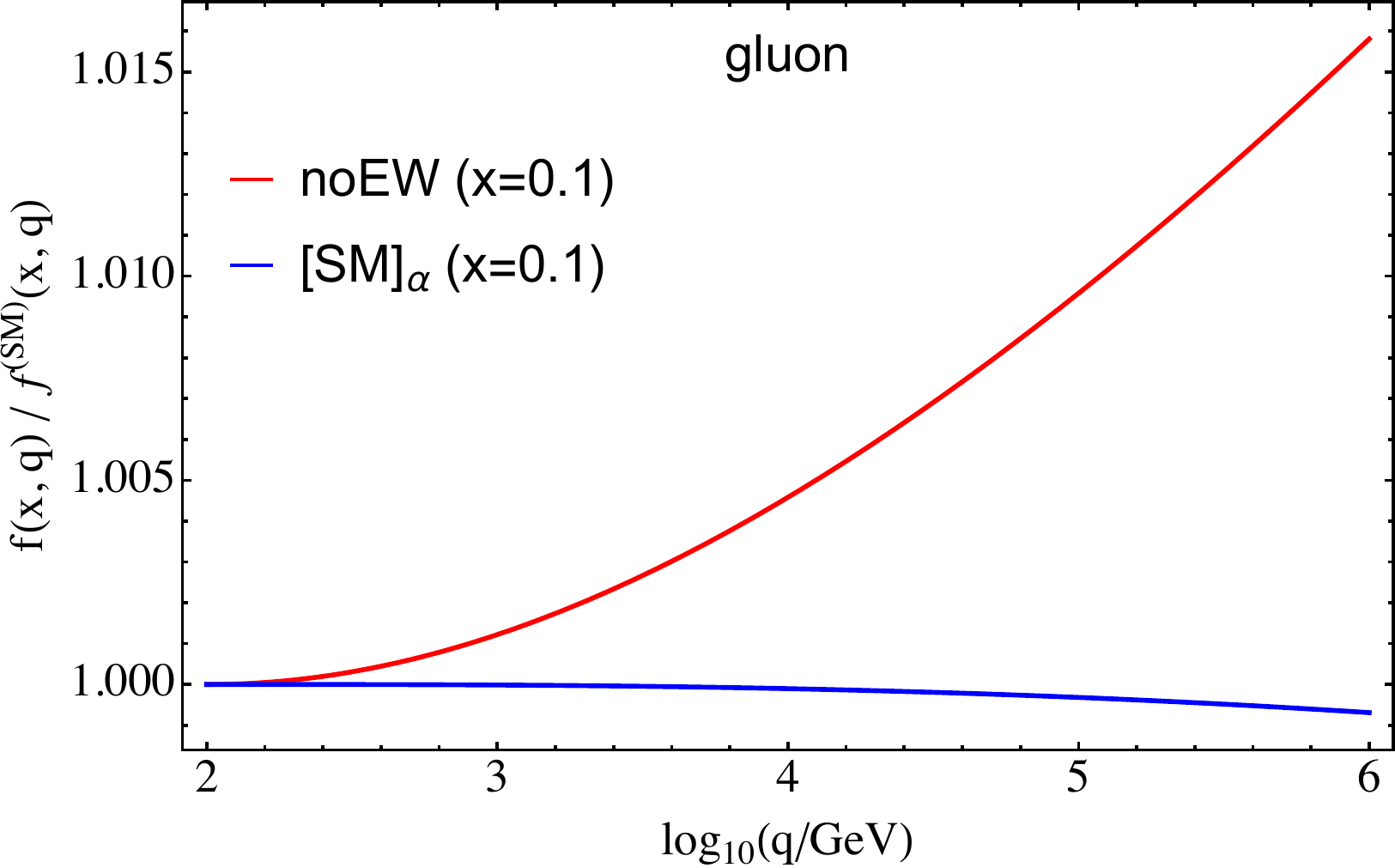}
  \includegraphics[scale=0.4]{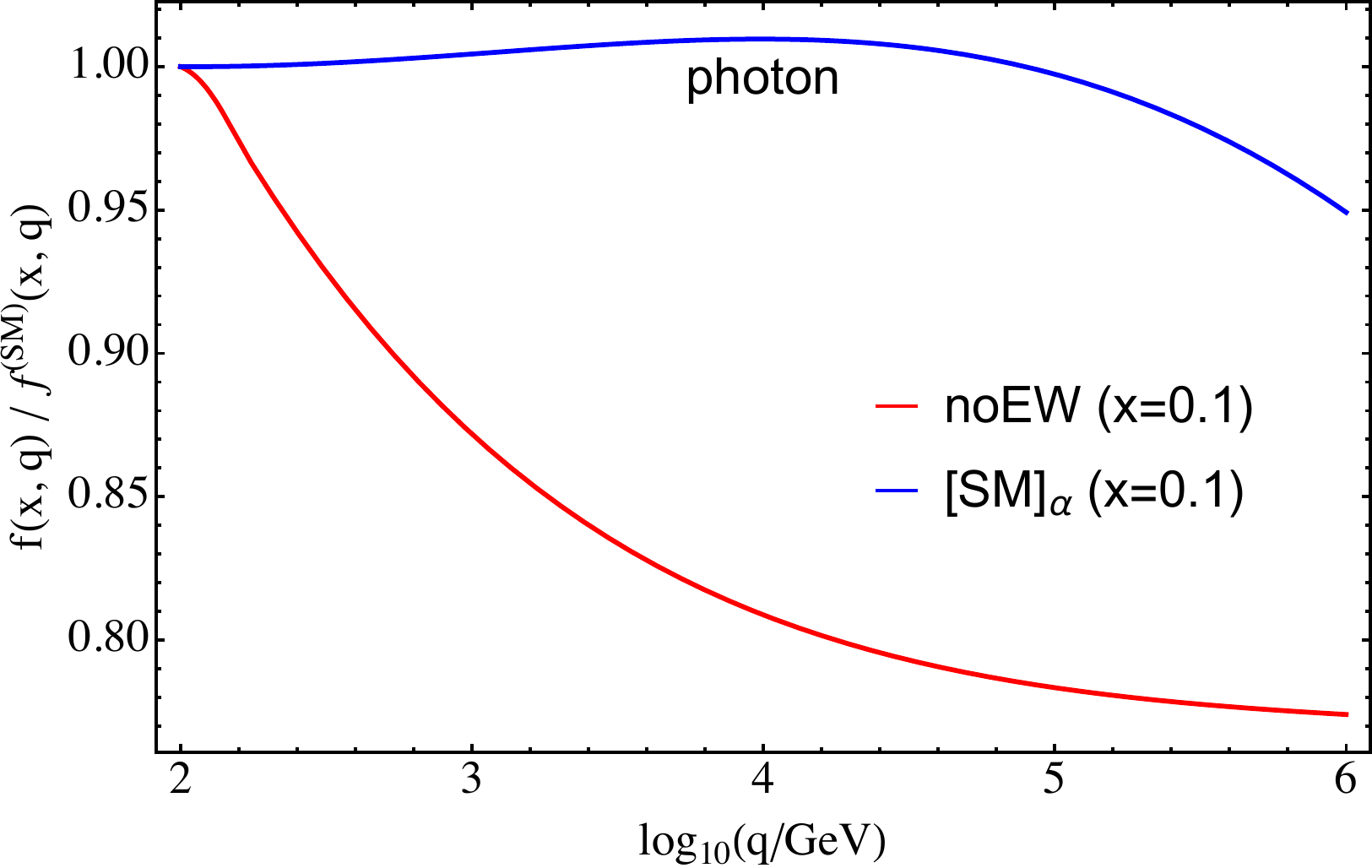}
\caption{\label{fig:PDFExpansion_V0}%
The ratio of the ``noEW'' and expanded SM PDFs relative to the PDF evaluated in the full SM for the massless vector bosons.  
}}
\FIGURE[h]{
 \centering
  \includegraphics[scale=0.4]{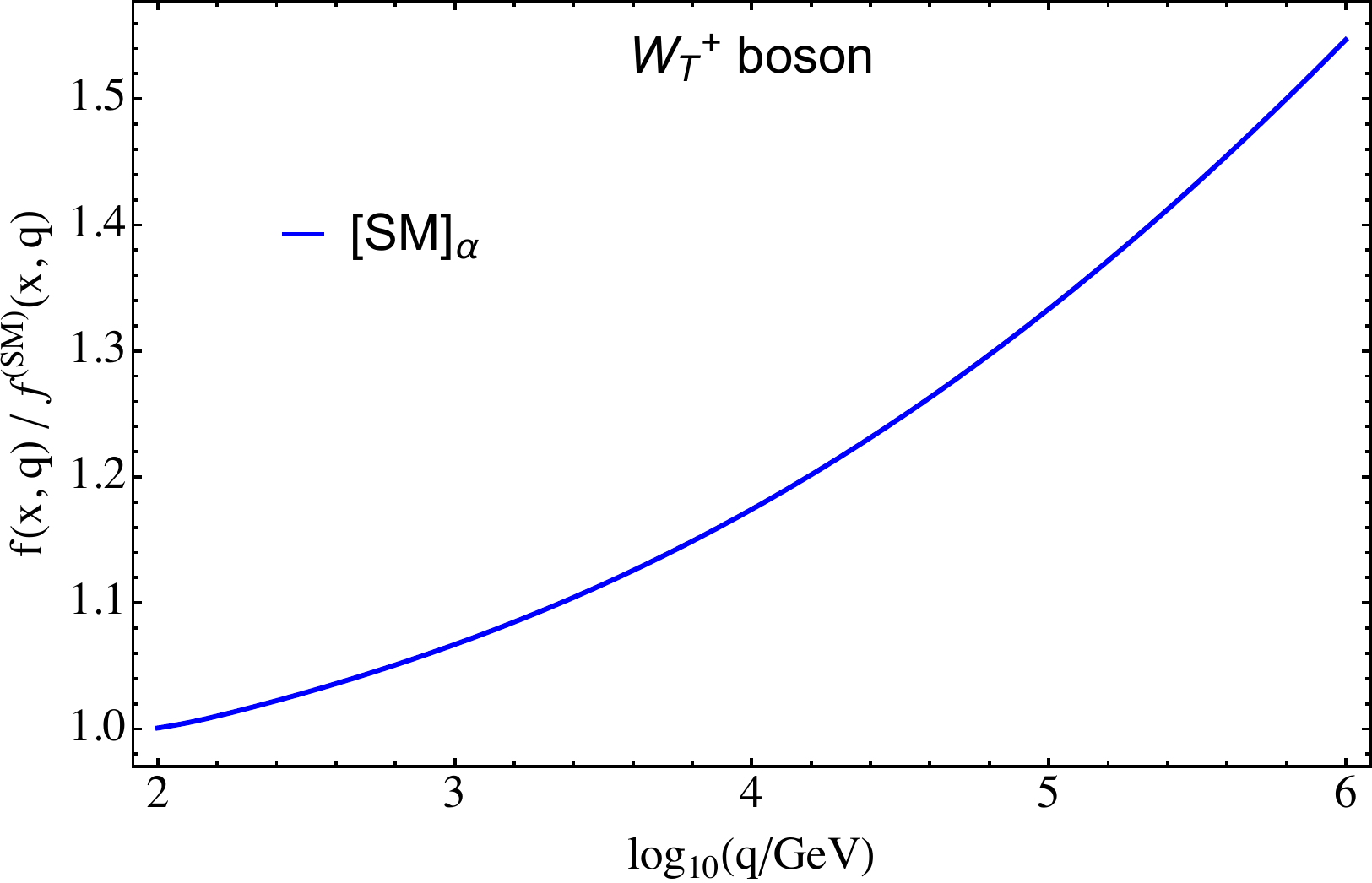}
  \includegraphics[scale=0.4]{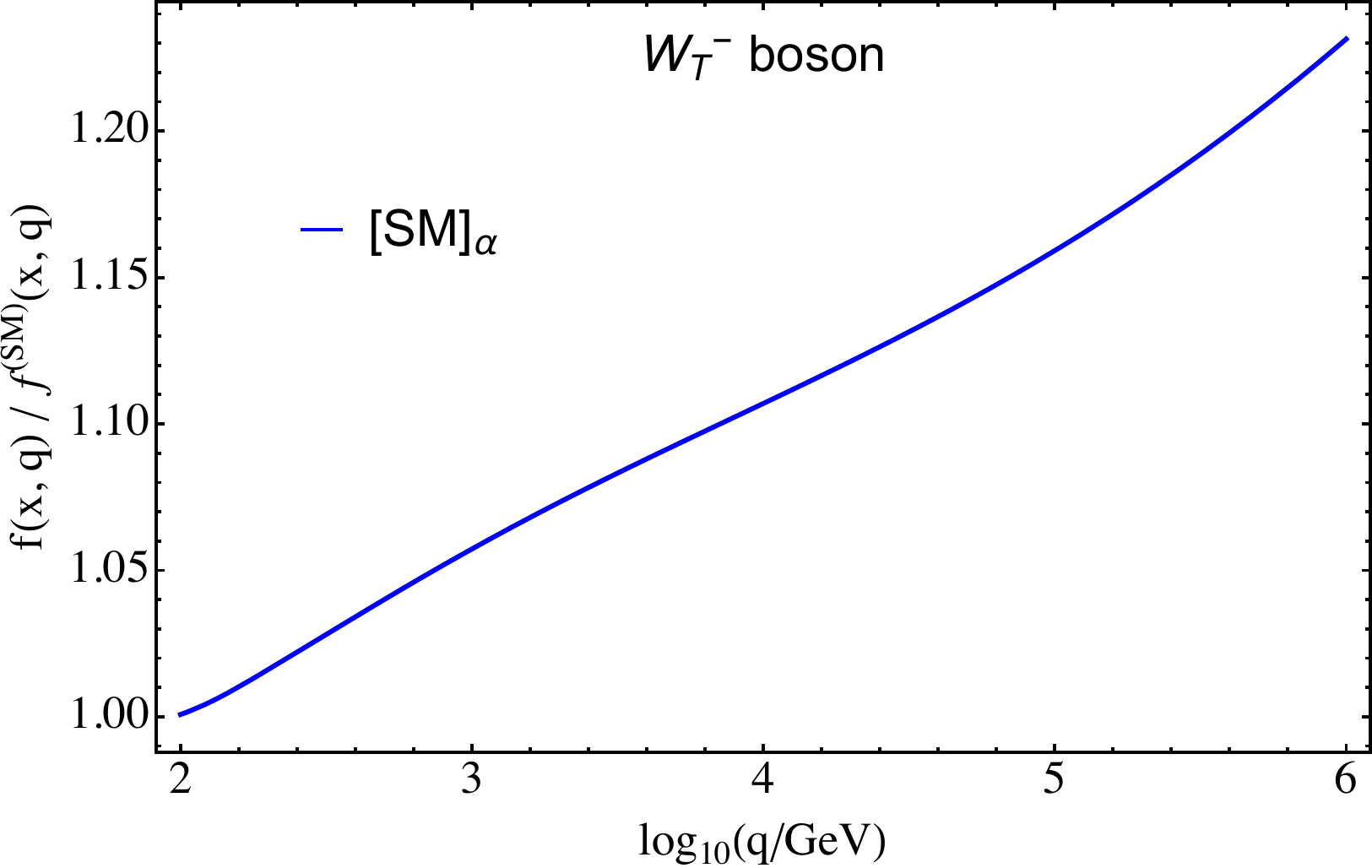}
  \includegraphics[scale=0.4]{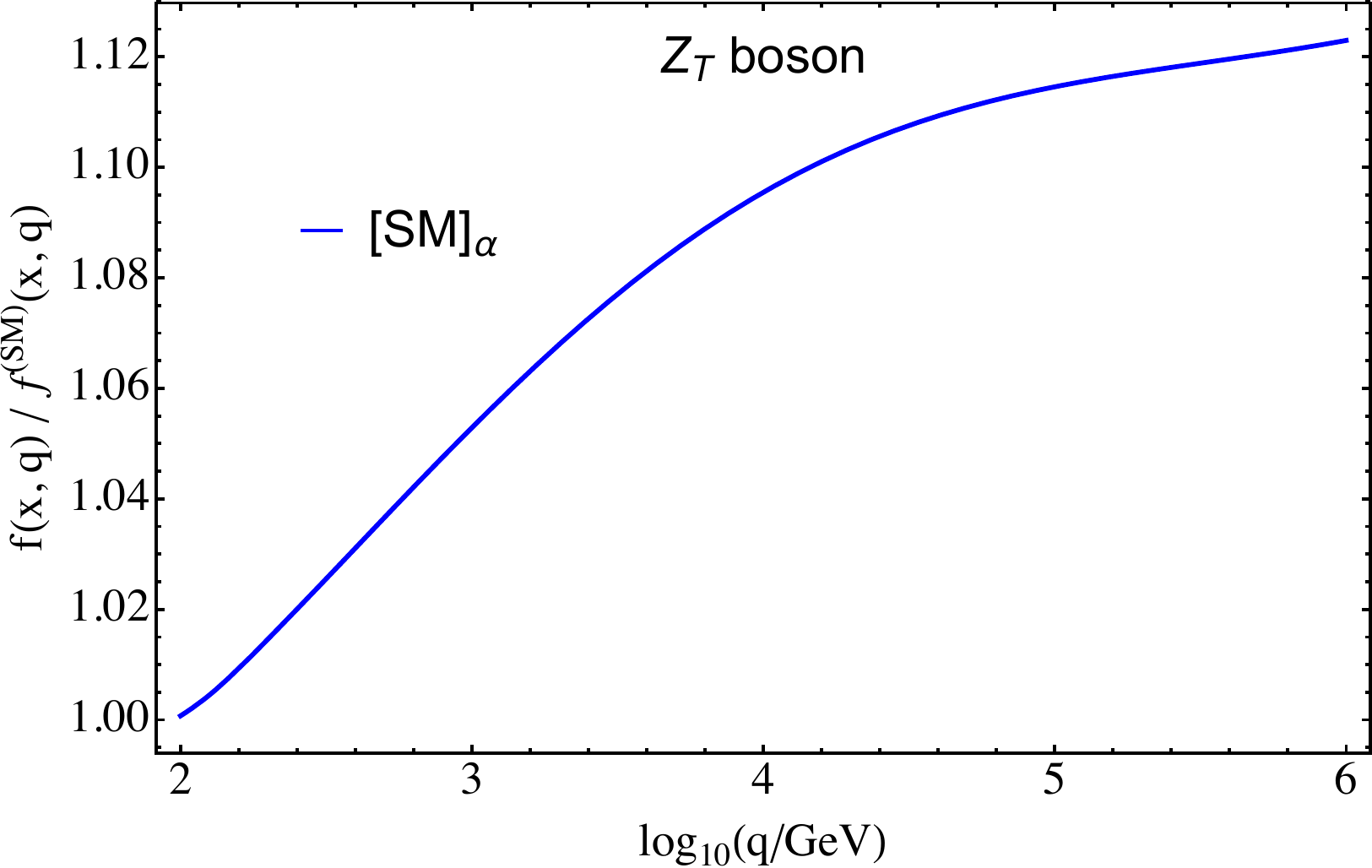}
  \includegraphics[scale=0.4]{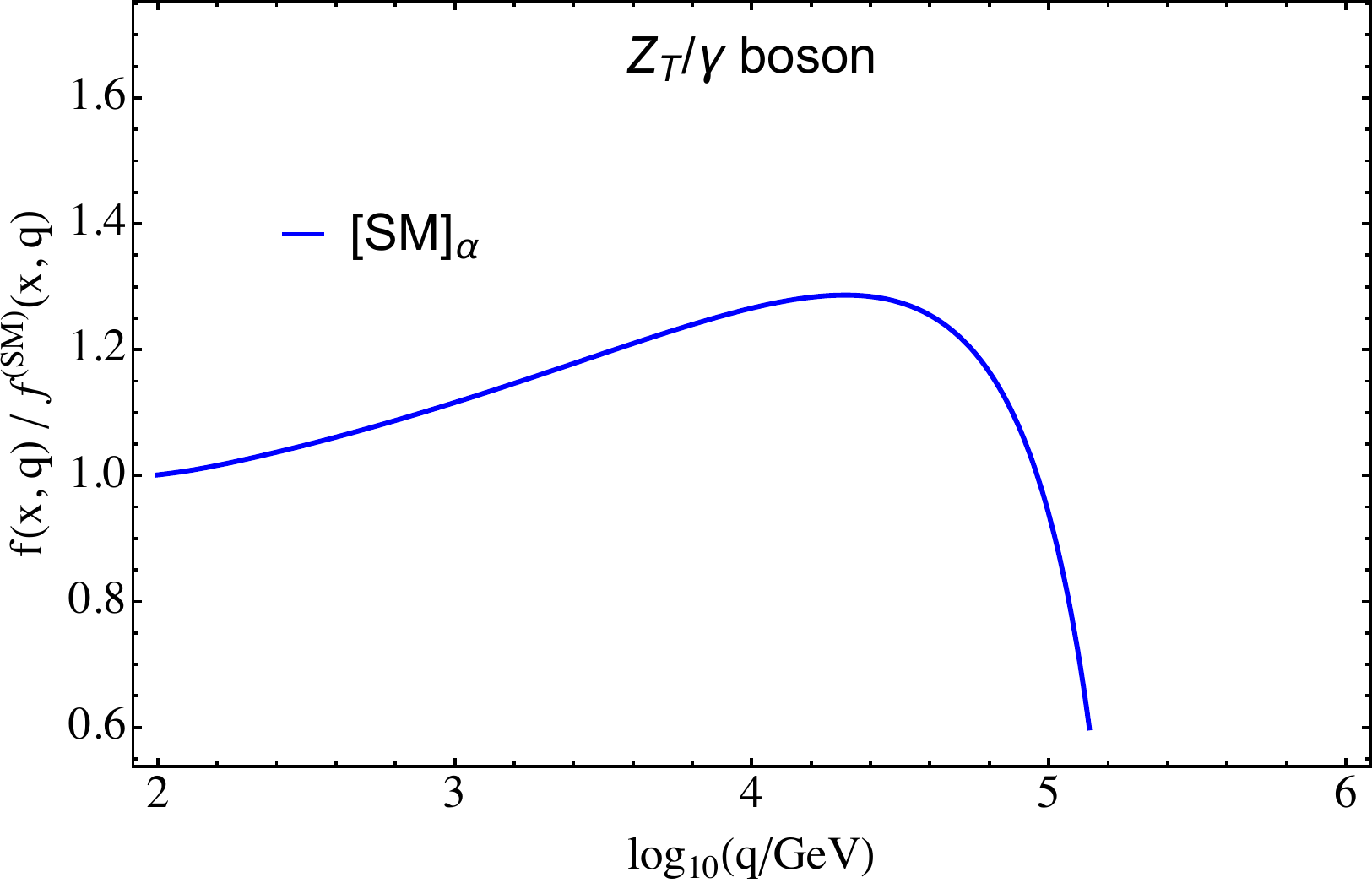}
\caption{\label{fig:PDFExpansion_VM}%
The ratio of the ``noEW'' and expanded SM PDFs relative to the PDF
evaluated in the full SM for the transversely-polarized massive vector bosons.  
}}
The results are shown in Fig.~\ref{fig:PDFExpansion_Q} for left-handed
up and down (anti)quarks. One can clearly see that at low values of
$q$ the second-order correction is much smaller than the first-order
correction, which is indicative of an absence of large logarithmic
corrections. For $q \gtrsim 10^4$ GeV, however, the logarithmic
contributions become noticeable, and the second-order correction grows
relative to the first order, becoming comparable to the latter, at
least for some of the PDFs, by $q \sim 10^6$ GeV.  Notice that
at high $q$ the left-handed up and down quarks move in opposite directions,
to restore isospin symmetry asymptotically.

For the gluon and photon, the results are shown in
Fig.~\ref{fig:PDFExpansion_V0}.  The gluon does not couple to the
massive vector bosons directly, so the electroweak effect is strongly
suppressed.  Since the ``noEW'' PDFs include only QCD evolution, the
photon does not evolve at all in that case, and receives a
large first-order EW correction.  In higher orders it can couple directly to the
massive bosons, so its PDF is double-logarithmically sensitive to the
ratio $m_V / q$.  Therefore, although the higher-order corrections are
much smaller than the first order for $q \sim q_V$, they
grow much more rapidly at high values of $q$. 

For massive vector bosons $r^{\rm noEW}(x,q)$ is zero, since their
PDFs vanish when only QCD effects are included for $q>q_V$. Therefore, given our
results, the validity of the perturbative expansion can only be
studied through the ratio $r^{\rm SM, \alpha}(x,q)$, whose deviation from
unity starts at first order in $\alpha_I$ as given
in~\eq{ratio_VM}. In Fig.~\ref{fig:PDFExpansion_VM} one sees clearly
the poor convergence of the perturbative expansion of massive boson
PDFs: the deviation from unity is much larger than one power of
$\alpha_I$, which of course is due to the double-logarithmic
dependence on $m_V / q$.  The ratio
between the expanded PDF and the full PDF can deviate from unity
by an amount in excess of 10\%. 

Given these PDFs and their expansions, one can find the first-order
expansion of the SM luminosity
\begin{align}
\label{eq:LumiSMExpanded}
\left[ {\cal L}^{\rm SM}_{AB}(x_A, x_B; Q)\right]_\alpha &= f_A^{\rm noEW}(x_A, Q) \, f_B^{\rm noEW}(x_B, Q) + f_A^{\rm noEW}(x_A, Q) \, g_B(x_B, Q)
\nn
 & \quad + g_A(x_A, Q) \, f_B^{\rm noEW}(x_B, Q) 
\,.
\end{align}
From the definition \eq{gdef} this obviously satisfies
\begin{align}\label{eq:LumiDiff}
{\cal L}^{\rm SM}_{AB}(x_A, x_B; Q) - \left[ {\cal L}^{\rm SM}_{AB}(x_A, x_B; Q)\right]_\alpha = {\cal O}(\alpha_I^2)
\,.\end{align}
Thus the difference in \eq{LumiDiff} can be used to add
resummation terms to a NLO calculation, since it excludes all terms in 
the luminosity ${\cal L}^{\rm SM}_{AB}$ at ${\cal O}(1)$ and ${\cal
  O}(\alpha_I)$ while including all LL terms of higher order.

Parton luminosities involving two massive gauge bosons (such as ${\cal
  L}_{ZZ}$, ${\cal L}_{W^+ W^-}$) only start to contribute at order
$\alpha_I^2$, since the PDF of each such boson is suppressed by one
power of $\alpha_I$. This means that their effect is not included in
the first-order expansion of the luminosity discussed above. However,
vector-boson fusion (VBF) processes (those involving two massive gauge bosons in the
initial state) can be significant numerically. For this reason, one might want to include
their effects exactly at lowest order, and only rely on the
LL approximation to predict their higher-order terms. This requires
subtraction of the ${\cal O}(\alpha_I^2)$ terms from ${\cal L}_{VV}$ when
computing the expanded luminosity. The resulting modified expanded luminosity
\begin{align}
\label{eq:LumiSMExpandedVfusions}
\left[ {\cal L}^{\rm SM}_{AB}(x_A, x_B; Q)\right]_{\alpha}^{\rm mod} &= f_A^{\rm noEW}(x_A, Q) \, f_B^{\rm noEW}(x_B, Q) + f_A^{\rm noEW}(x_A, Q) \, g_B(x_B, Q)
\nn
 & \quad + g_A(x_A, Q) \, f_B^{\rm noEW}(x_B, Q)  + g_A(x_A, Q) \, g_B(x_B, Q) \delta_{AB,VV}
\,,\end{align}
coinciding with \eq{LumiSMExpanded} for all channels except $V_TV_T$,
allows the inclusion of the exact lowest-order $V_TV_T$ contribution
together with all resummed higher-order terms in that and the other channels.
Thus to combine a fixed-order calculation including all EW effects at
NLO, as well as the VBF process $V_TV_T$ at LO, which we 
denote by
\begin{align}
\langle O \rangle_{\rm NLO+VV} \equiv \langle O \rangle_{\rm NLO} + \langle O \rangle^{\rm VV}_{\rm LO} 
\,,\end{align}
one would compute
\begin{align}
\label{eq:sigmaNLO_LL_Vfusion}
\langle O \rangle_{\rm NLO+VV+LL} = \langle O \rangle_{\rm NLO+VV} +
  \langle O \rangle_{\rm LL} - \left[ \langle O \rangle_{\rm
  LL}\right]^{\rm mod}_{\alpha}
\,.
\end{align}
where
\begin{align}
\left[\langle O\rangle_{\rm LL}\right]^{\rm mod}_{\alpha} = \sum_{AB}
  \int \! \df \Phi_n \, O_n(\Phi_n) \, \left[ {\cal L}^{\rm
  SM}_{AB}(x_A, x_B; Q)\right]_\alpha^{\rm mod} B_{AB}(\Phi_n)
\,.
\end{align}

\FIGURE[h]{
 \centering
  \includegraphics[scale=0.23]{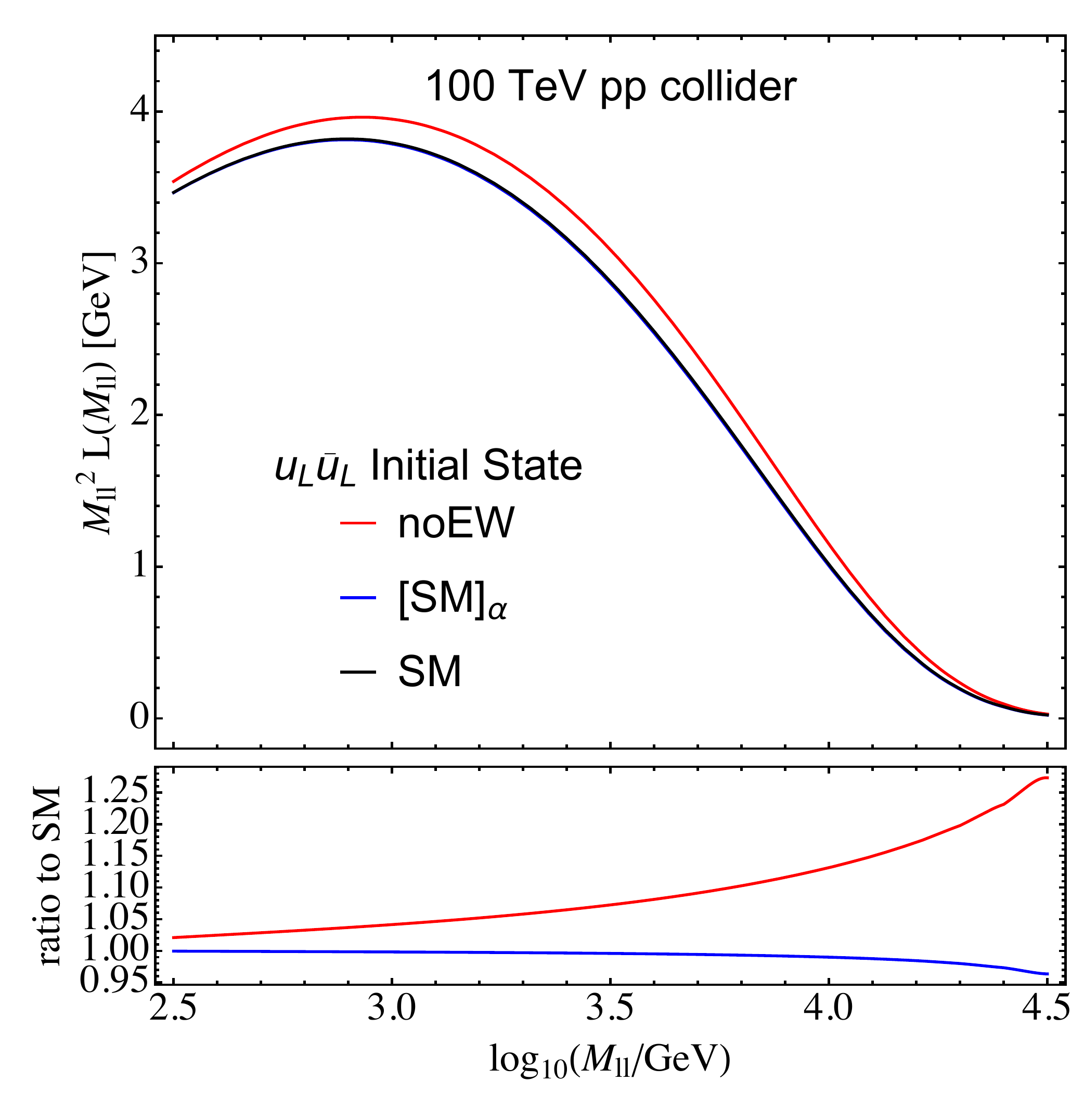}
  \includegraphics[scale=0.23]{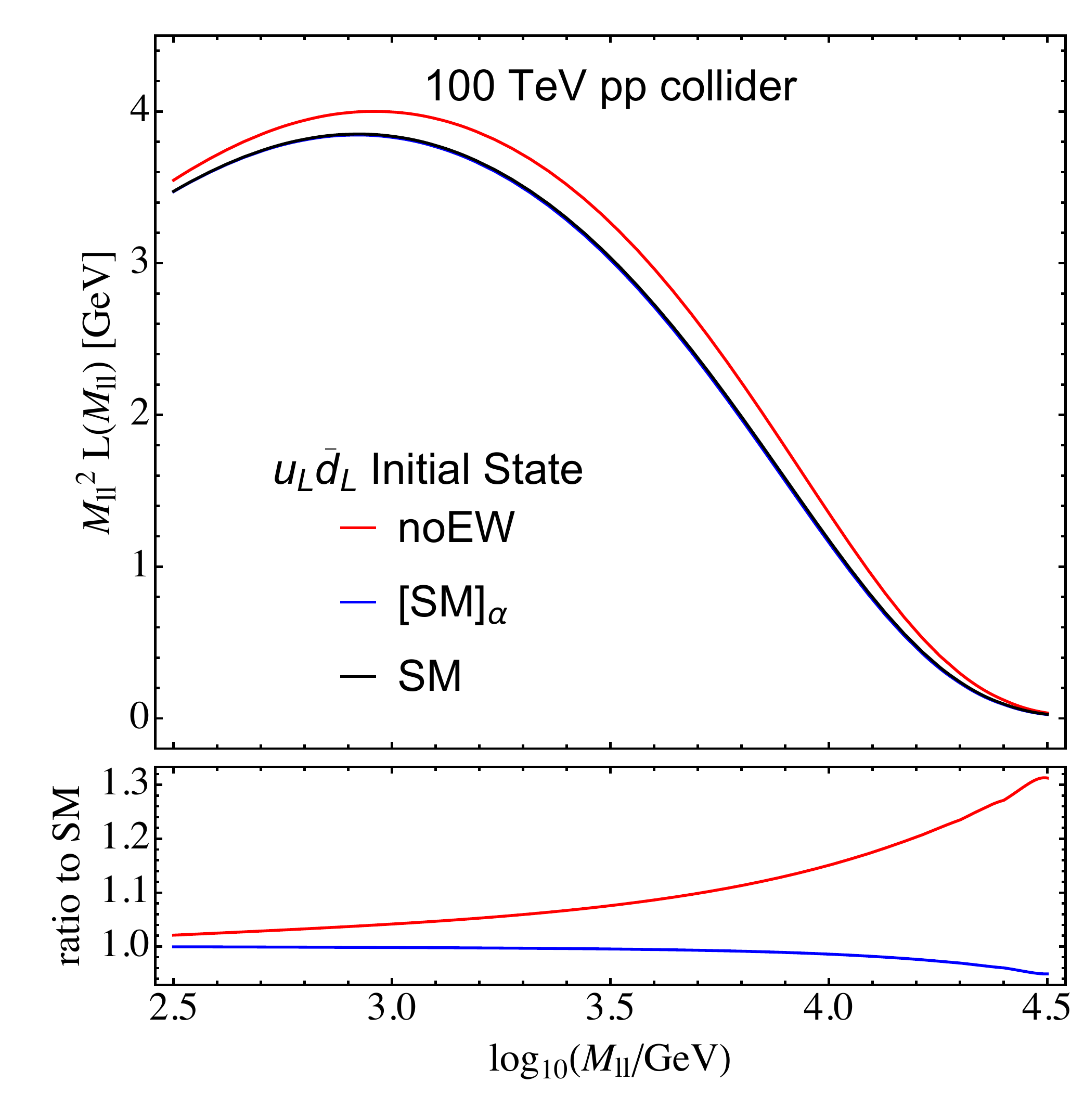}
  \includegraphics[scale=0.23]{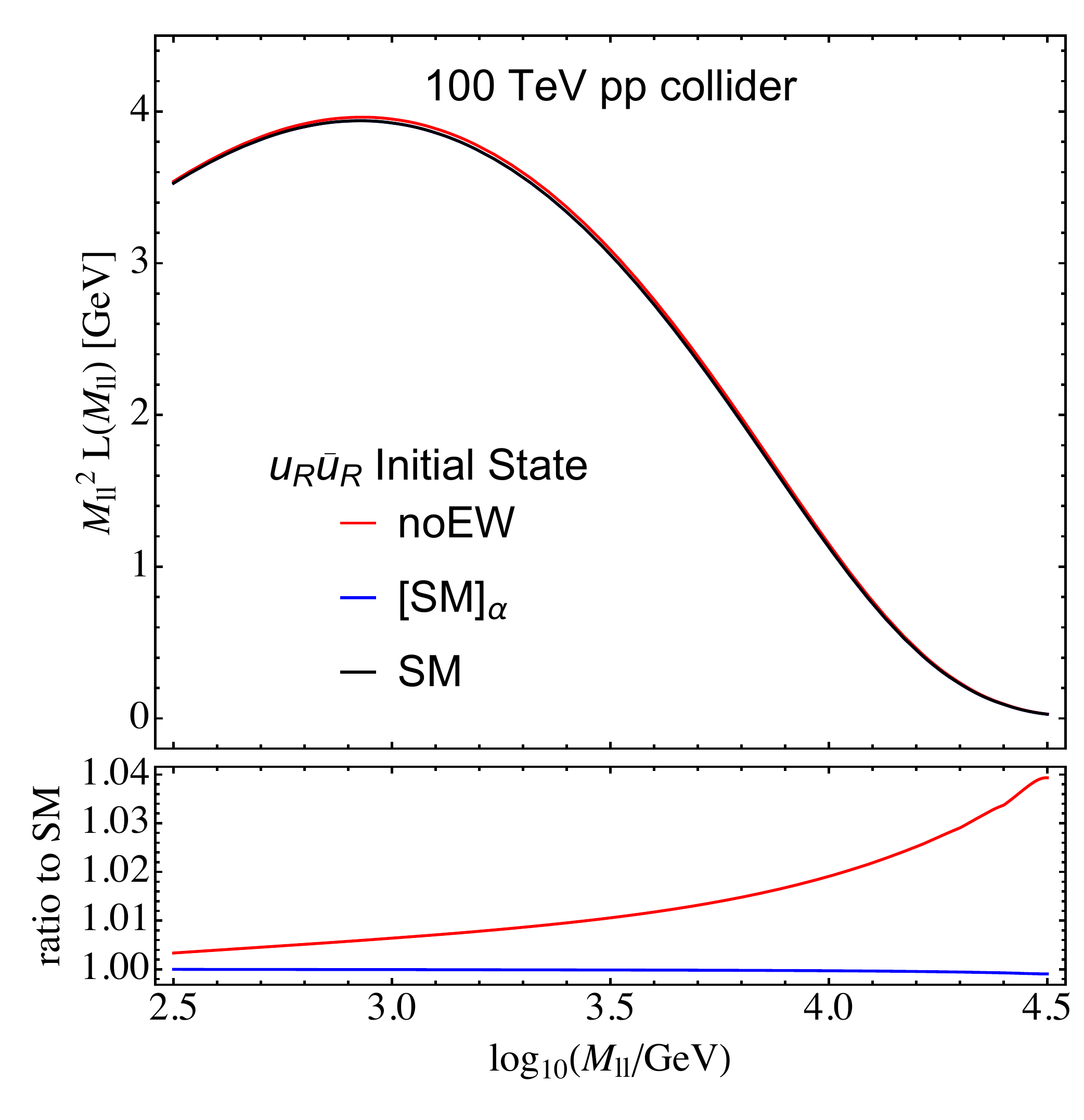}
  \includegraphics[scale=0.23]{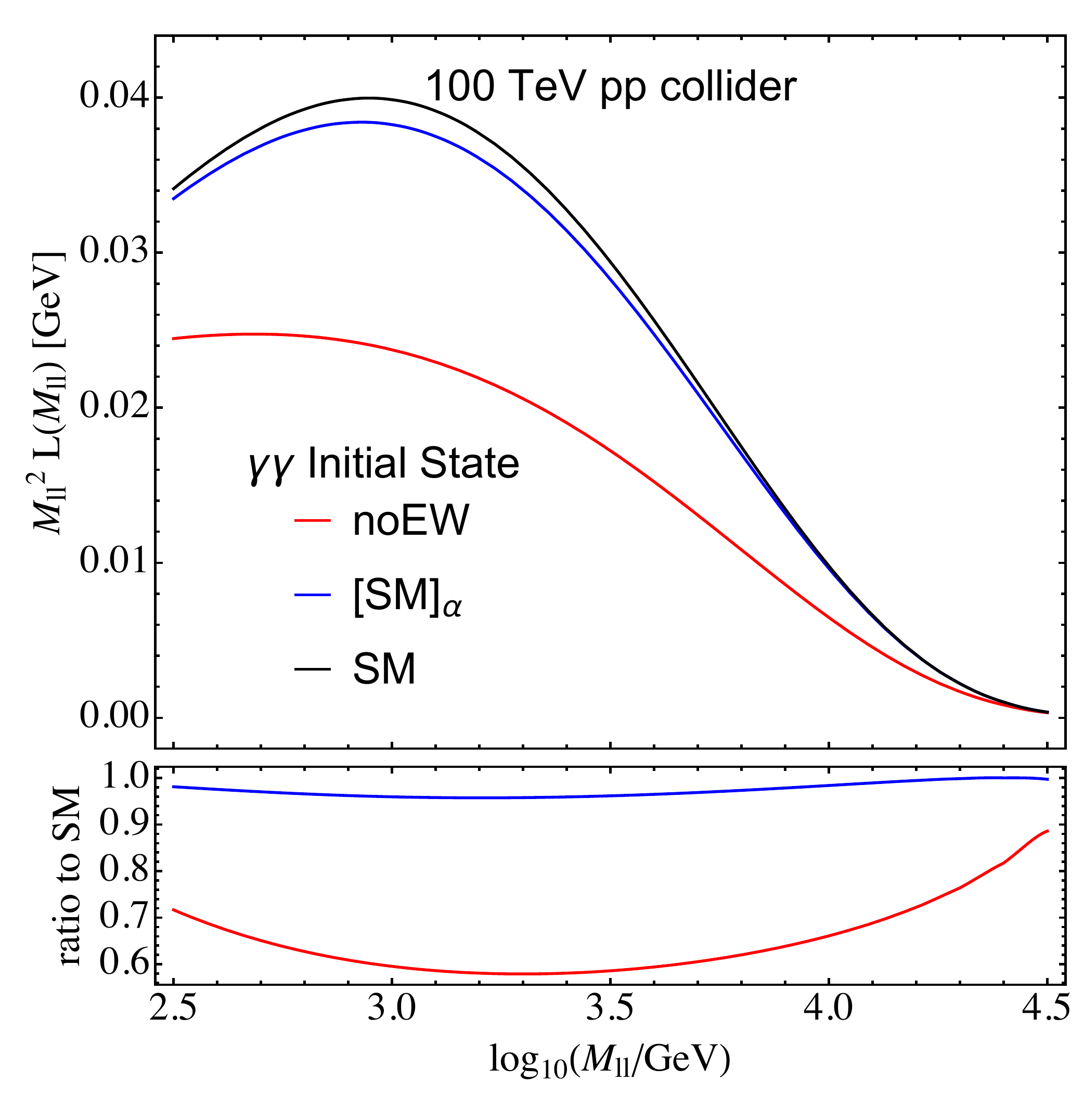}
  \includegraphics[scale=0.23]{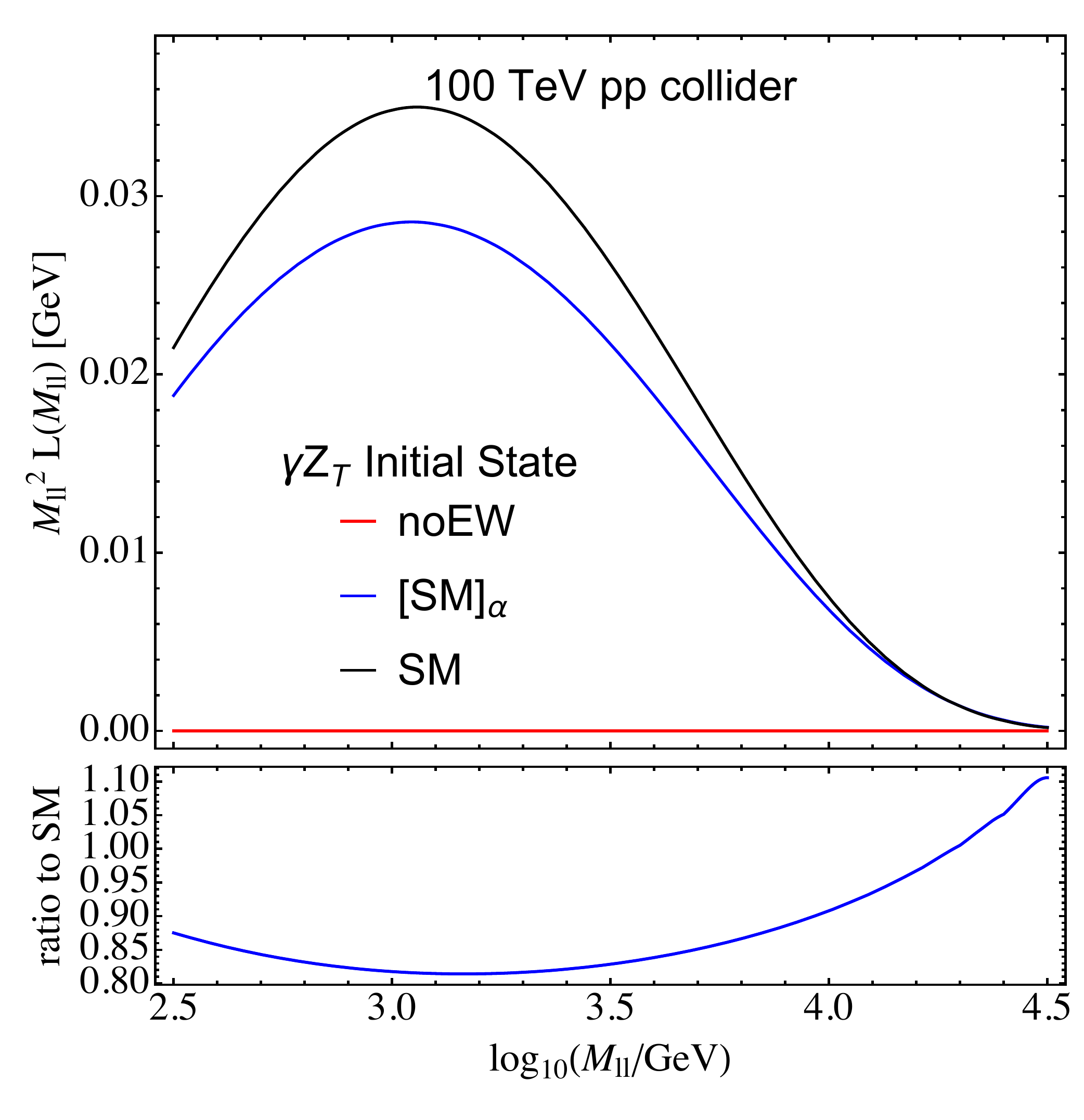}
  \includegraphics[scale=0.23]{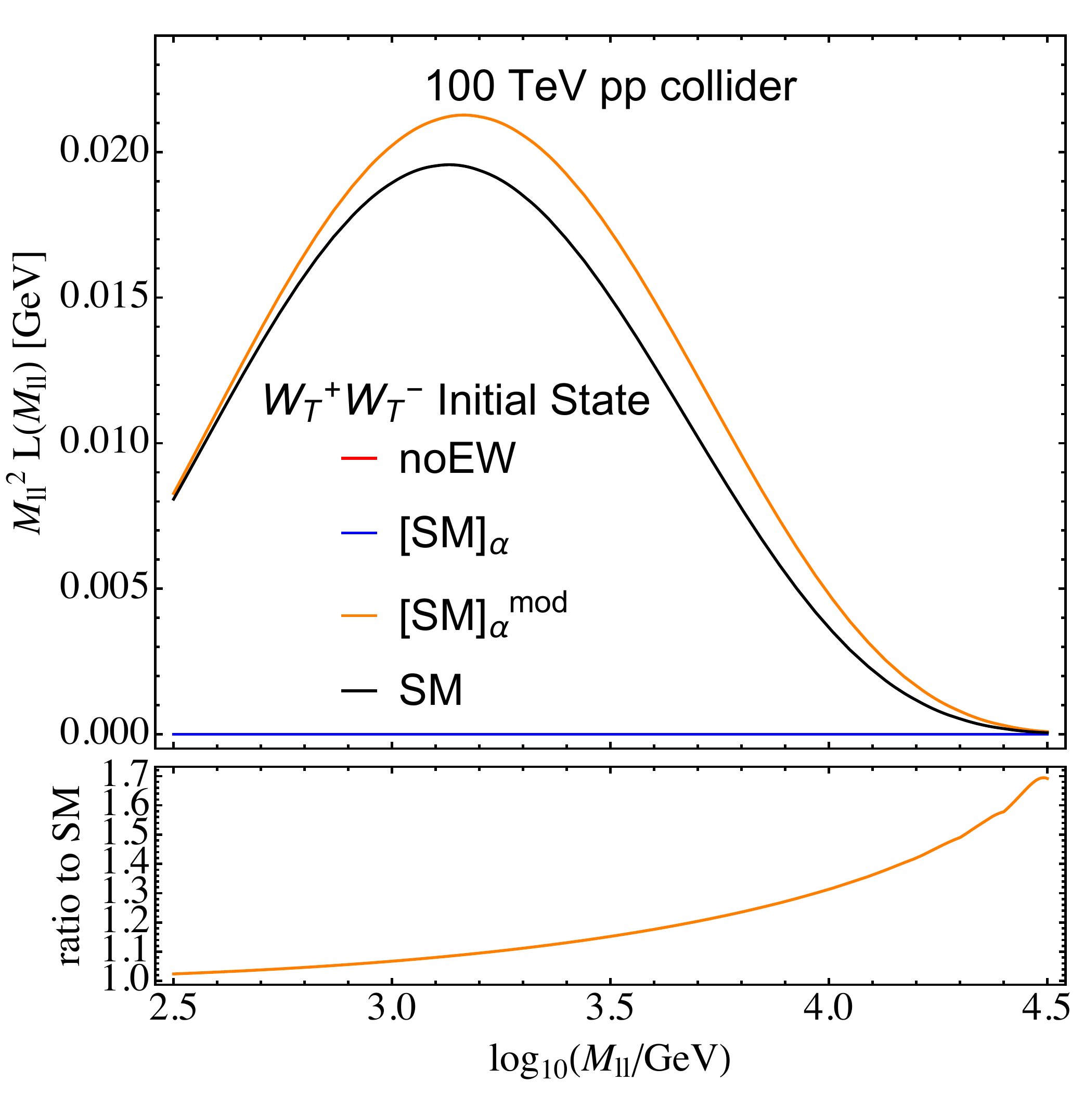}
\caption{\label{fig:lumi_100}%
Plots showing luminosities for various choices of initial states. We show in black ${\cal L}^{\rm SM}$, in red ${\cal L}^{\rm noEW}$, in blue $\left[{\cal L}^{\rm SM}\right]_\alpha$ and for $V_TV_T$ initial states in orange $\left[{\cal L}^{\rm SM}\right]_\alpha^{\rm mod}$.
}}

In Fig.~\ref{fig:lumi_100} we show the results for a few selected parton luminosities
\begin{align}
{\cal L_{AB}}(M_{\ell\ell}) = \int\! \df x_A\,\df x_B \, {\cal L}^{\rm
  SM}_{AB}\left(x_A, x_B; M_{\ell\ell}\right) \,\delta\left(M_{\ell\ell}-\sqrt{x_1x_2 S}\right)
\,,
\end{align}
for $pp$ collisions at $\sqrt S=100$ TeV,
rescaled by the square of the invariant mass $M_{\ell\ell}$ to
overcome the steeply falling nature of the functions. We show in black
${\cal L}^{\rm SM}$, in red ${\cal L}^{\rm noEW}$, in blue
$\left[{\cal L}^{\rm SM}\right]_\alpha$ and for VV initial states in
orange $\left[{\cal L}^{\rm SM}\right]_\alpha^{\rm mod}$. One can see
that for left-handed quarks the difference between ${\cal L}^{\rm
  noEW}$ and ${\cal L}^{\rm SM}$ is larger than the difference between
$\left[{\cal L}^{\rm SM}\right]_\alpha$ and ${\cal L}^{\rm SM}$ for
all values of $M_{\ell\ell}$ considered, indicating that the double
logarithms are not yet large enough to have
$\alpha\ln^2(M_{\ell\ell}^2/m_V^2) \gtrsim 1$.
However, for $M_{\ell\ell}\gtrsim$ a few TeV the higher-order
terms become significant. For right-handed quarks, there are no
double logarithms and the coupling is smaller, so the convergence of the
perturbation series is much faster. For the $\gamma\gamma$ initial state,
recall that the ``noEW'' photon PDF is frozen at the matching scale
$q_V=$ 100 GeV so the order $\alpha$ correction is large
and dominates the expansion. For the
$\gamma Z$ luminosity the higher-order terms are more
significant.  Finally, the $W^+ W^-$ luminosity
vanishes for $\left[{\cal L}^{\rm SM}\right]_\alpha$. Using the
modified expansion reproduces the dominant features of the full
luminosity, but higher order terms are still very important for
$M_{\ell \ell} \gtrsim$ few TeV. 

\section{Resummation of logarithms in inclusive di-lepton production}
\label{sec:diLeptonProduction}

In this section, we study the effects of higher-order leading
logarithms in the process of  fully inclusive di-lepton production.
This will allow us to assess the correction from logarithmic
resummation that needs to be applied to fixed-order calculations
in order to achieve NLO+LL accuracy.
Note, however, that we do not include the fixed-order calculation here.

The definition of  fully inclusive di-lepton production
was given in the introduction, but we will repeat it here for
completeness. The inclusive process is defined to include a
lepton-antilepton pair of any charge
of a given generation and any number of extra gauge bosons in the
final state. So to NLO EW accuracy, this process sums over the final
states $\ell^+ \ell^- (+V)$, $\ell^+ \nu_\ell (+V)$, $\bar \nu_\ell
\ell^- (+V)$, $\bar \nu_\ell \nu_\ell (+V)$. Here $\ell$ denotes, for
example, the electron and $\nu_\ell$ the electron neutrino and the
$(+V)$ denotes the possible addition of a $\gamma$, $Z$ or $W^{\pm}$
boson. Since we are summing over both electrons and neutrinos, and we
are including the radiation of extra electroweak gauge bosons,
the final state of this process is SU(2) symmetric, as required. In order to 
regulate the strong enhancement of forward lepton production in vector
boson fusion, we impose a cut on the transverse momentum 
of each lepton $p_T > 100$ GeV. This implies that the accessible di-lepton
invariant masses are $M_{\ell\ell}> 200$ GeV.

\FIGURE[h]{
 \centering
  \includegraphics[scale=0.5]{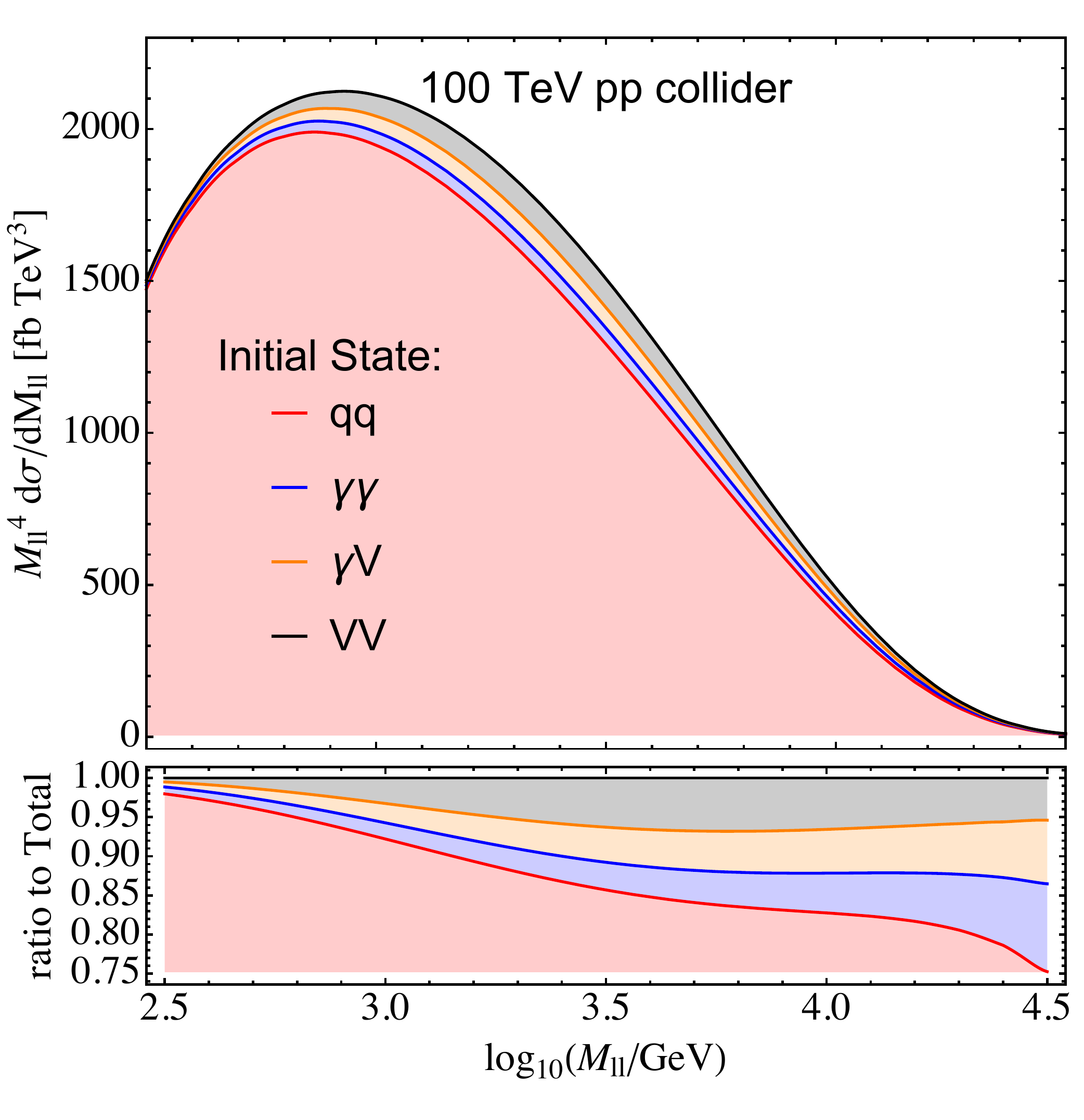}
\caption{\label{fig:dSigmadM_Stacked_100}%
The differential cross section $M_{\ell\ell}^4 \df \sigma / \df M_{\ell\ell}(p_{T\ell}>100$ GeV) for a 100 TeV collider, showing the makeup of the total cross section in terms of the individual initial states.
}}

To compute the partonic Born cross section $B_{AB}(\widehat \Phi_n)$
in \eq{sigmaLL}, one relates it to the square of the relevant matrix element via
\begin{align}
\label{eq:BornDef}
B_{AB}(\Phi_n) \equiv \frac{1}{4p_A \mcdot p_B} \left| M(AB\to\ell\ell)\right|^2
\,,\end{align}
where $\ell$ denotes either a charged lepton or a neutrino. As
discussed in \sect{introduction}, for the initial states $A$ and $B$
one needs $q \bar q$ of all possible quark flavors and helicities, as well as $V V$,
where $V$ can be any one of the electroweak gauge bosons, $\gamma,
Z^0, W^\pm,$ or the mixed $\gamma/Z^0$ representing interference
contributions.  The contributions of initial-state leptons,
longitudinal gauge bosons and Higgs bosons are much smaller and will
be neglected.  Details of the cross-section calculations are given in
the Appendix. 

The leading-logarithmic differential cross section $\df \sigma / \df M_{\ell\ell}$
is shown for a 100 TeV $pp$ collider in
Fig~\ref{fig:dSigmadM_Stacked_100}.  In order to make the plot easier
to read, we have multiplied the differential cross section by
$M_{\ell\ell}^4$ to overcome its steeply falling nature. We have
stacked the contributions of the various initial states $q \bar q$, 
$\gamma\gamma$, $\gamma V_T$ and $V_TV_T$ (where $V_T$ now denotes a
sum over massive transversely polarized electroweak gauge bosons) on top of each
other. In the lower part of the plot, we show the ratio to the total
contribution, giving a better estimate of the relative size of each
contribution. One can see that the dominant contribution is from the
$q \bar q$ initial states, but the relative size of the initial states
with two vector bosons grows with increasing $M_{\ell\ell}$. For a 100
TeV collider, the contributions with vector bosons in the initial
state are around 25\% for $M_{\ell\ell} = 10^{4.5}$ GeV $\sim 30$ TeV.
 
\FIGURE[h]{
	\centering
	\includegraphics[scale=0.26]{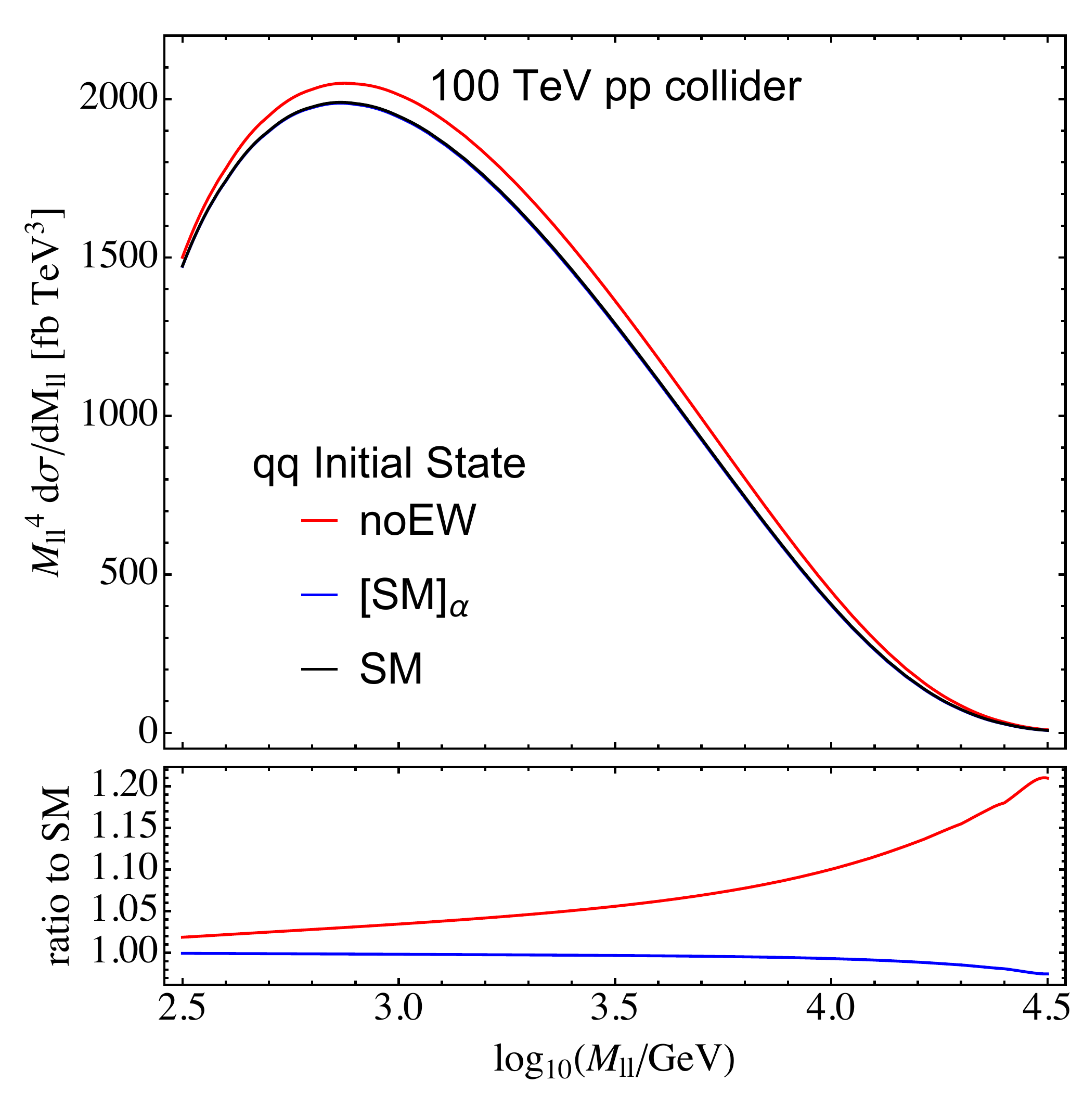}
	\includegraphics[scale=0.26]{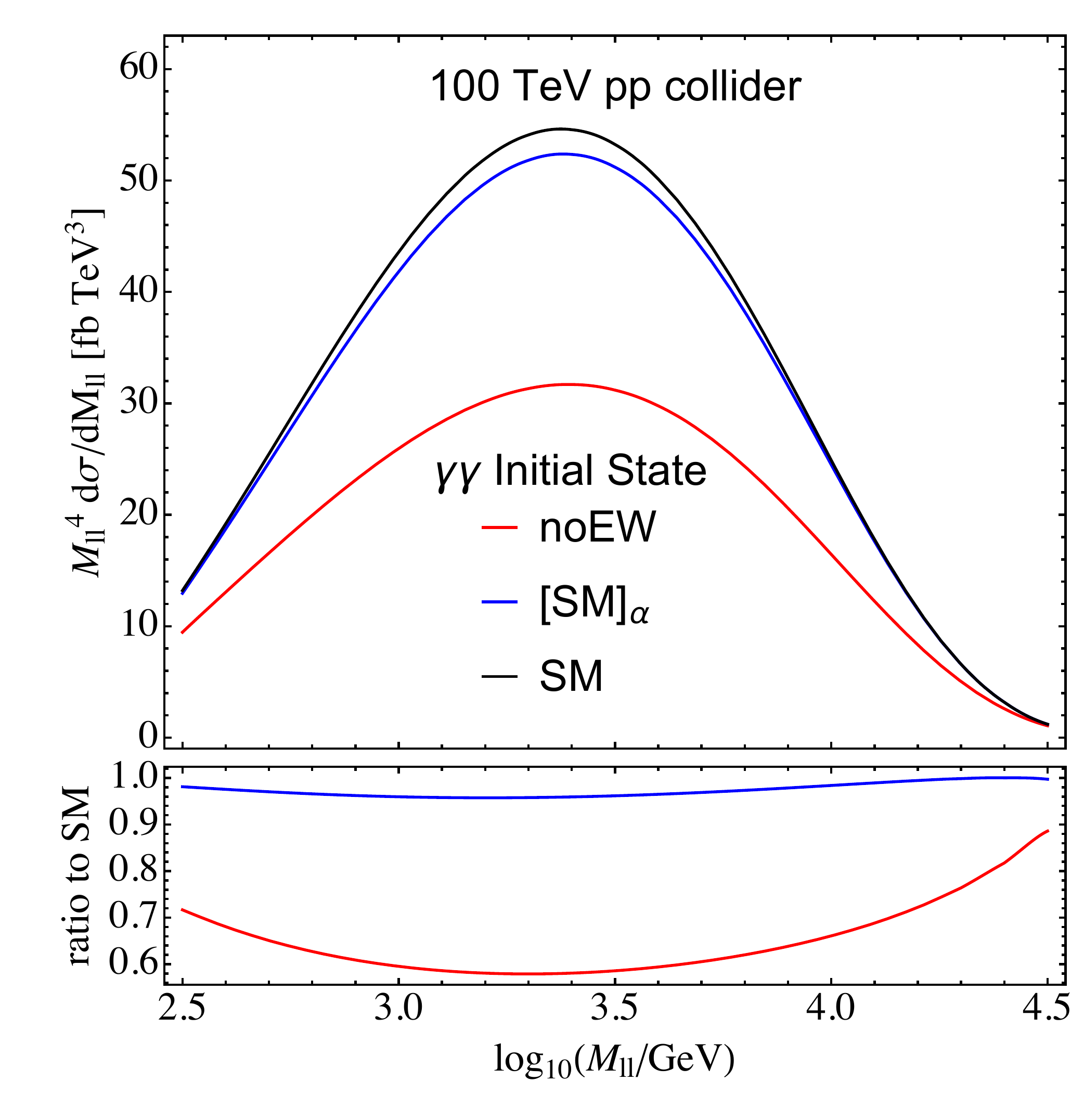}
	\includegraphics[scale=0.26]{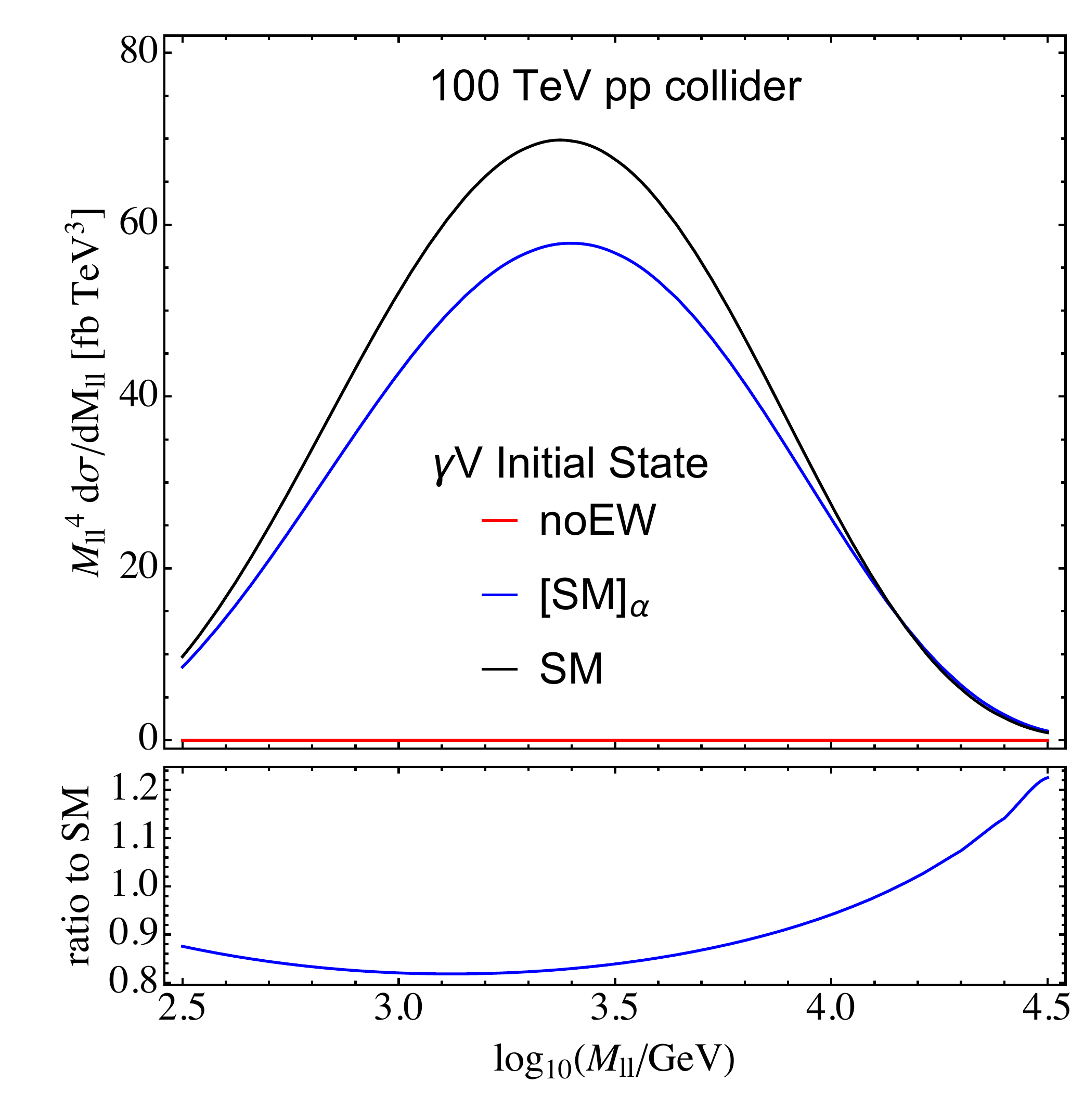}
	\includegraphics[scale=0.26]{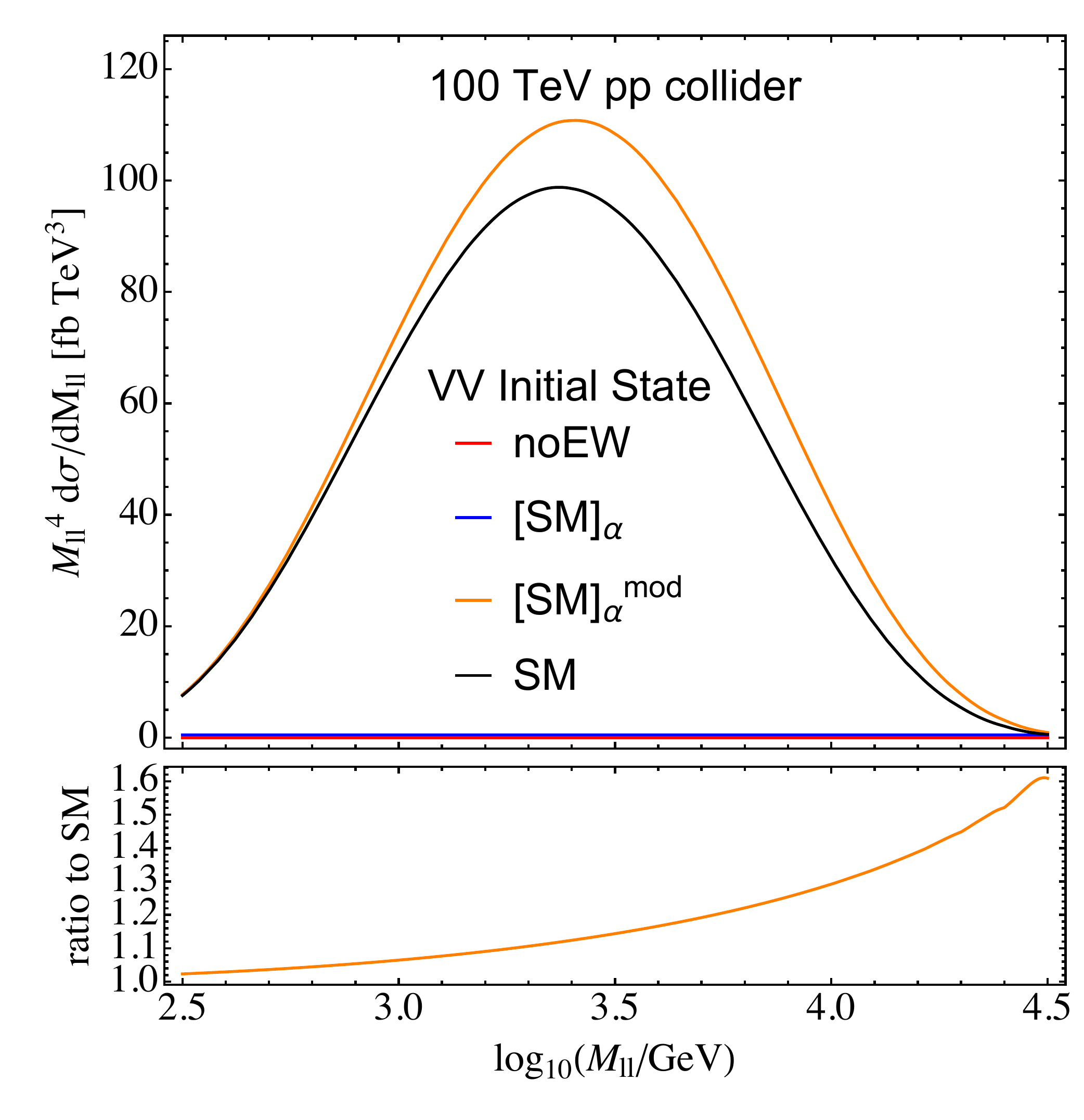}
	\caption{\label{fig:convergence_100}%
		The expansion of the various contributions to
                $M_{\ell\ell}^4 \df \sigma / \df
                M_{\ell\ell}(p_{T\ell}>100$ GeV) for a 100 TeV collider. We show in black the result obtained using ${\cal L}^{\rm SM}$, in red that using ${\cal L}^{\rm noEW}$, in blue that using $\left[{\cal L}^{\rm SM}\right]_\alpha$ and for $V_TV_T$ initial states in orange that using $\left[{\cal L}^{\rm SM}\right]_\alpha^{\rm mod}$.
}}
 
Next, we take each of the four contributions and investigate their
convergence under EW perturbation theory. For this, we compare the
result of the LL-resummed cross section $M_{\ell\ell}^4 \df \sigma_{\rm
  LL} / \df M_{\ell\ell}$ with its first-order expansion $[M_{\ell\ell}^4\df
\sigma_{\rm LL} / \df M_{\ell\ell}]_\alpha$ for the various initial
states. The results are shown for a 100 TeV pp collider in
Fig.~\ref{fig:convergence_100}, where in black we show the resummed result,
and in blue its first-order expansion. The difference between these
two is the correction that should be added to a fixed-order
calculation to achieve NLO+LL accuracy.  For comparison, we also show
in red the ``noEW'' result. The
difference between the blue and red curves shows the logarithmically
enhanced order-$\alpha$ contribution. As one can
see, for the $q\bar q$ channel, the expansion of the LL result is
quite close to the full LL result, indicating that the higher-order
corrections are quite small. This is due to two facts: First, the
right-handed quarks do not receive any double-logarithmic contributions
(and their single logarithmic terms come with coupling constant $\alpha_1$
rather than $\alpha_2$). Second, since sea quarks are mostly
iso-singlet, the double logarithms only arise from iso-vector
contributions of the valence quarks. Each of these facts reduces the
double logarithmic effect by roughly a factor of 2, such that overall the
effect is smaller by a factor of 4 compared to an individual $q_L \bar
q_L$ parton luminosity. Note that one of these factors of two would be
absent for a $p \bar p$ collider, so one would expect the
effect to be larger there by a factor of 2.

For $\gamma\gamma$ initial states, one needs to keep in mind that our
definition of $f^{\rm noEW}$ does not include any QED evolution for $q
> q_V$. This means that the photon PDF freezes out at the scale $q_V$
for this PDF. Since the effect of the evolution is of the same size as
the value of the PDF at $q = q_V$, the first order (difference of red
and black) gives an ${\cal O}(1)$ effect. The second order (difference
of blue and black) is considerably smaller than the first order for
all values of $M_{\ell\ell}$, but from the absolute value of the
correction it is also clear that the expansion parameter is much
larger than $\alpha_{\rm em} / \pi$ as one might naively expect. For
example, for $M_{\ell\ell} \sim 1$ TeV, the second-order correction
is almost 5\%.

Any process with massive bosons in the initial states is suppressed by
one power of $\alpha$ for each.  Therefore the ``noEW''  luminosity
vanishes for $\gamma V_T$ and $V_T V_T$,  and for $V_T V_T$ the
[SM]$_\alpha$ luminosity also vanishes,  as indicated by
the red and blue lines in the last two
plots. However, for $\gamma V_T$ the second-order correction (the difference
between the blue and the black line) reaches tens of percent at high
$M_{\ell\ell}$, indicating that the higher order perturbative
corrections are significant. For $V_TV_T$ initial states, we also show in orange the result
of the modified expansion \eq{LumiSMExpandedVfusions}, which includes
the leading ${\cal  O}(\alpha^2)$ term. The difference between the orange and black
curve denotes ${\cal O}(\alpha^3)$ terms, which are tens of percent of the leading 
${\cal O}(\alpha^2)$ terms, indicating again a poorly convergent perturbation series.

\FIGURE[h]{
	\centering
	\includegraphics[scale=0.4]{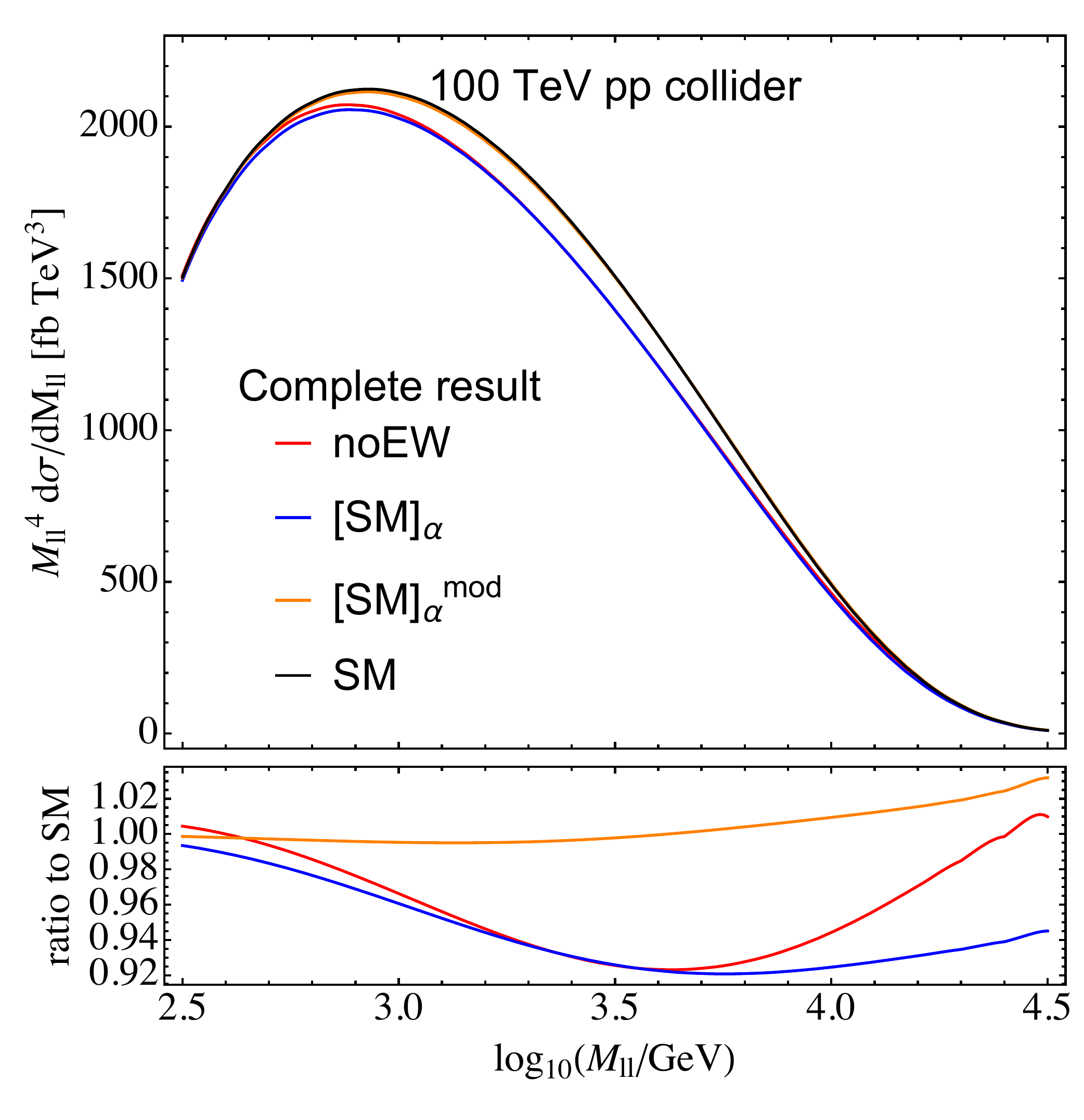}
	\caption{\label{fig:total_convergence}%
		The expansion of the complete result $M_{\ell\ell}^4 \df \sigma / \df M_{\ell\ell}(p_{T\ell}>100$ GeV) for a 100 TeV collider. The colors are the same as in Fig.~\ref{fig:convergence_100}.
}}

Putting these results together, we show in
Fig.~\ref{fig:total_convergence} the combination of the various
channels. One can see that perturbation theory is not very well
behaved and for $M_{\ell\ell} \gtrsim 5$ TeV, the second order
correction is essentially of the same size as the first order
correction (there is an accidental cancellation for very large
$M_{\ell\ell}$ which makes the first order correction become
small). The overall effect of the corrections of order $\alpha_I^2$
and higher for $M_{\ell\ell} \gtrsim$ a few TeV is of the order of
5\%. Most of this comes from the VBF processes, so the correction
to the modified expansion \eq{LumiSMExpandedVfusions} is much smaller.

To understand how these results depend on the center-of-mass energy of
the collider, we also show results at  27 TeV, which is the energy
that might be achieved by a high-energy upgrade of the LHC using
novel magnet technology~\cite{Tommasini:2017gkw}, and a
fictitious 1 PeV collider. In Fig.~\ref{fig:dSigmadM_Stacked_other}
the relative importance of the various channels is shown. One obvious
effect is that at high energies one has access to larger values of the
di-lepton invariant mass, for which the logarithmic enhancement is
stronger.  However, even at fixed invariant mass the 
relative importance of the initial states with vector
bosons is diminished (enhanced) for a 27 TeV (1 PeV) collider. This is
because at higher energies one is probing smaller values of $x$, and
the vector boson PDFs, like that of the gluon, rise rapidly with
decreasing $x$.  For a 1 PeV collider at the highest accessible di-lepton
invariant mass, the contribution of vector boson initial states is
almost 50\% of the total cross section.

Finally, we study the convergence of perturbation theory for individual channels for a 27 TeV and 1 PeV collider in Figs.~\ref{fig:convergence_27} and \ref{fig:convergence_1000}, respectively, and the complete result in Fig.~\ref{fig:total_convergence_other}. Qualitatively the effects are the same as for a 100 TeV collider, but the overall size of the effects are decreased (increased) for the 27 TeV (1 PeV) collider. 
\FIGURE[h]{
	\centering
	\includegraphics[scale=0.26]{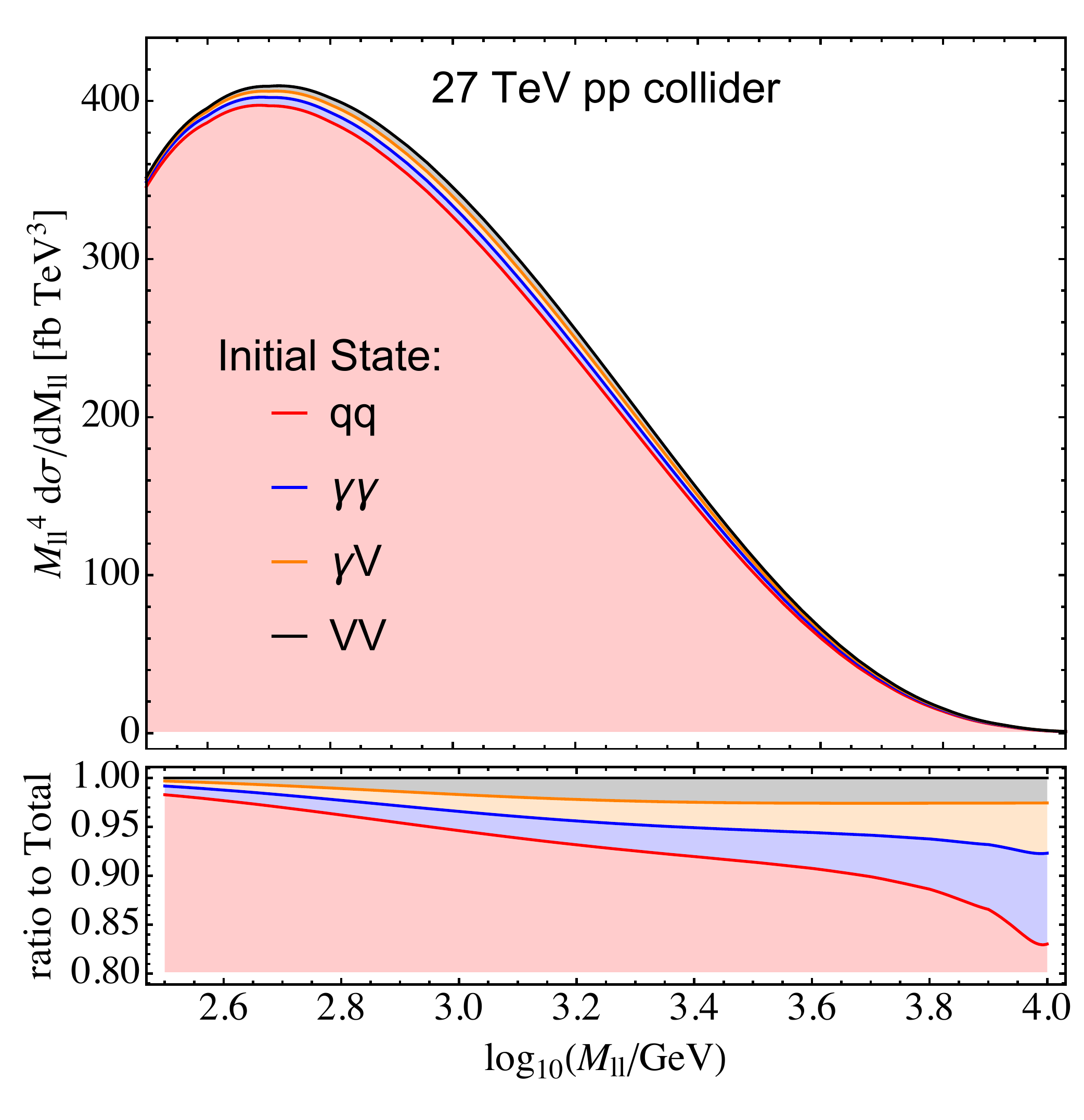}
	\includegraphics[scale=0.26]{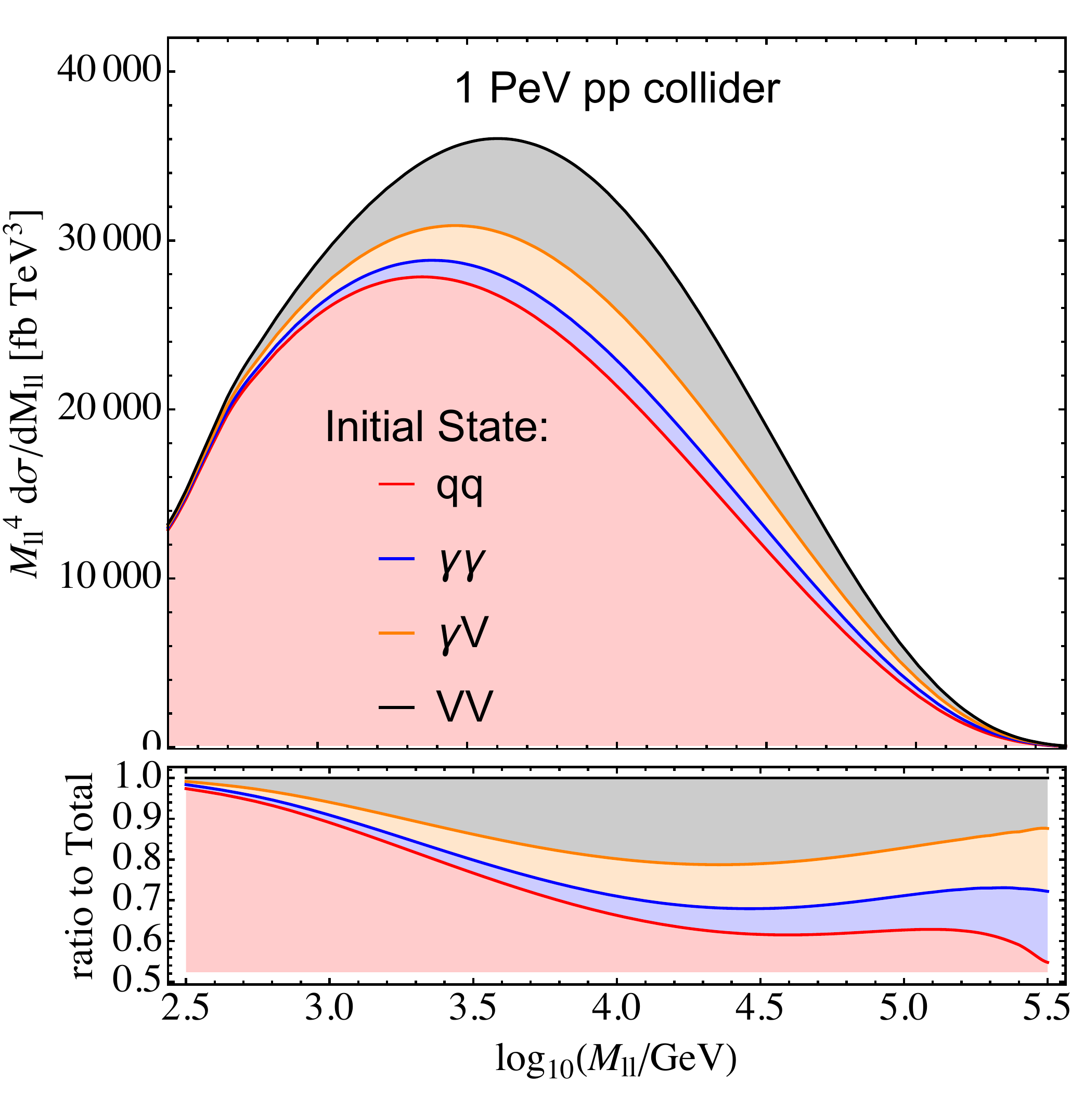}
	\caption{\label{fig:dSigmadM_Stacked_other}%
		The differential cross section $M_{\ell\ell}^4 \df \sigma / \df M_{\ell\ell}(p_{T\ell}>100$ GeV) for a $27$ TeV and $1$ PeV collider, showing the makeup of the total cross section in terms of the individual initial states.
}}

\FIGURE[h]{
	\centering
	\includegraphics[scale=0.26]{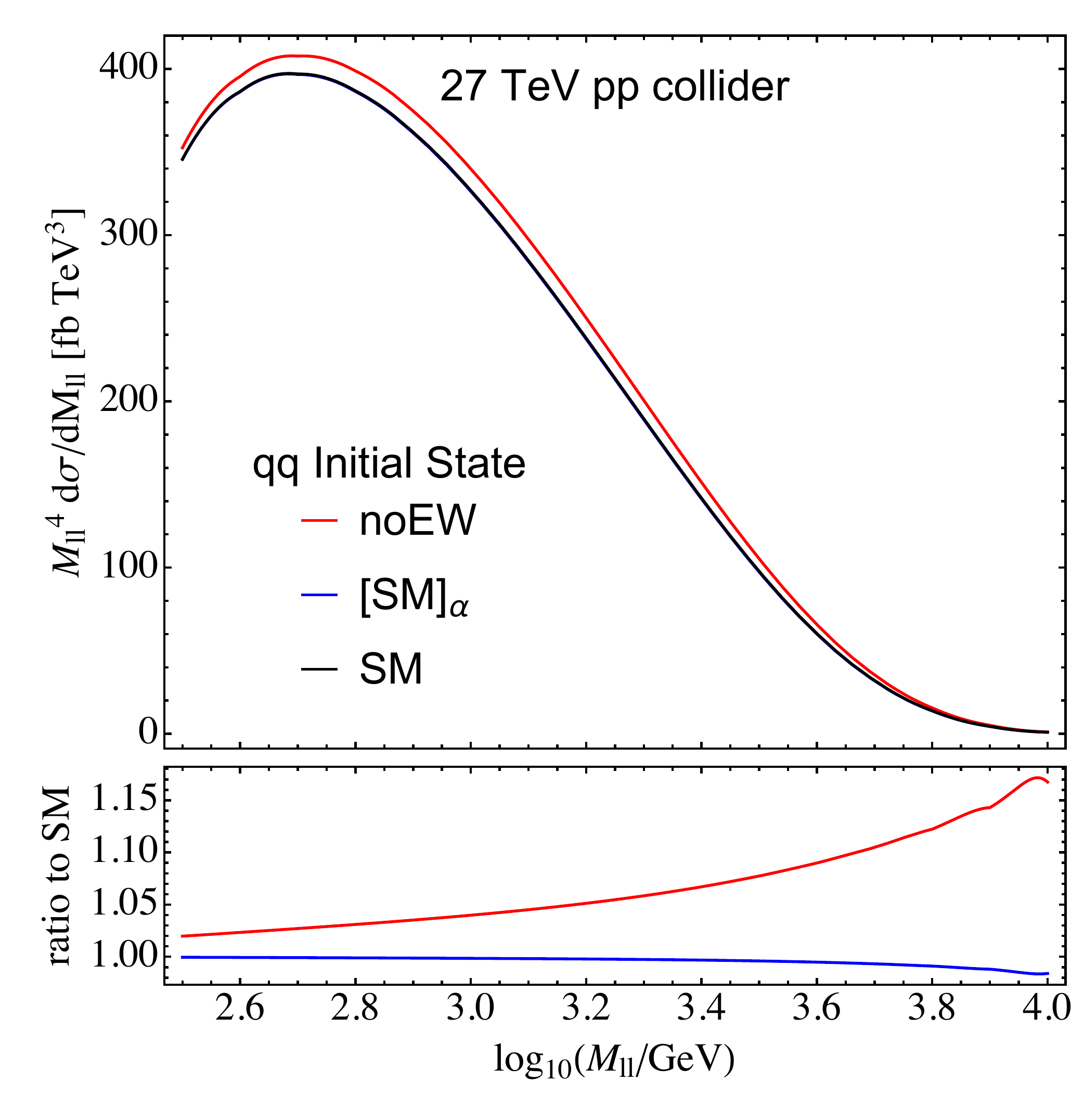}
	\includegraphics[scale=0.26]{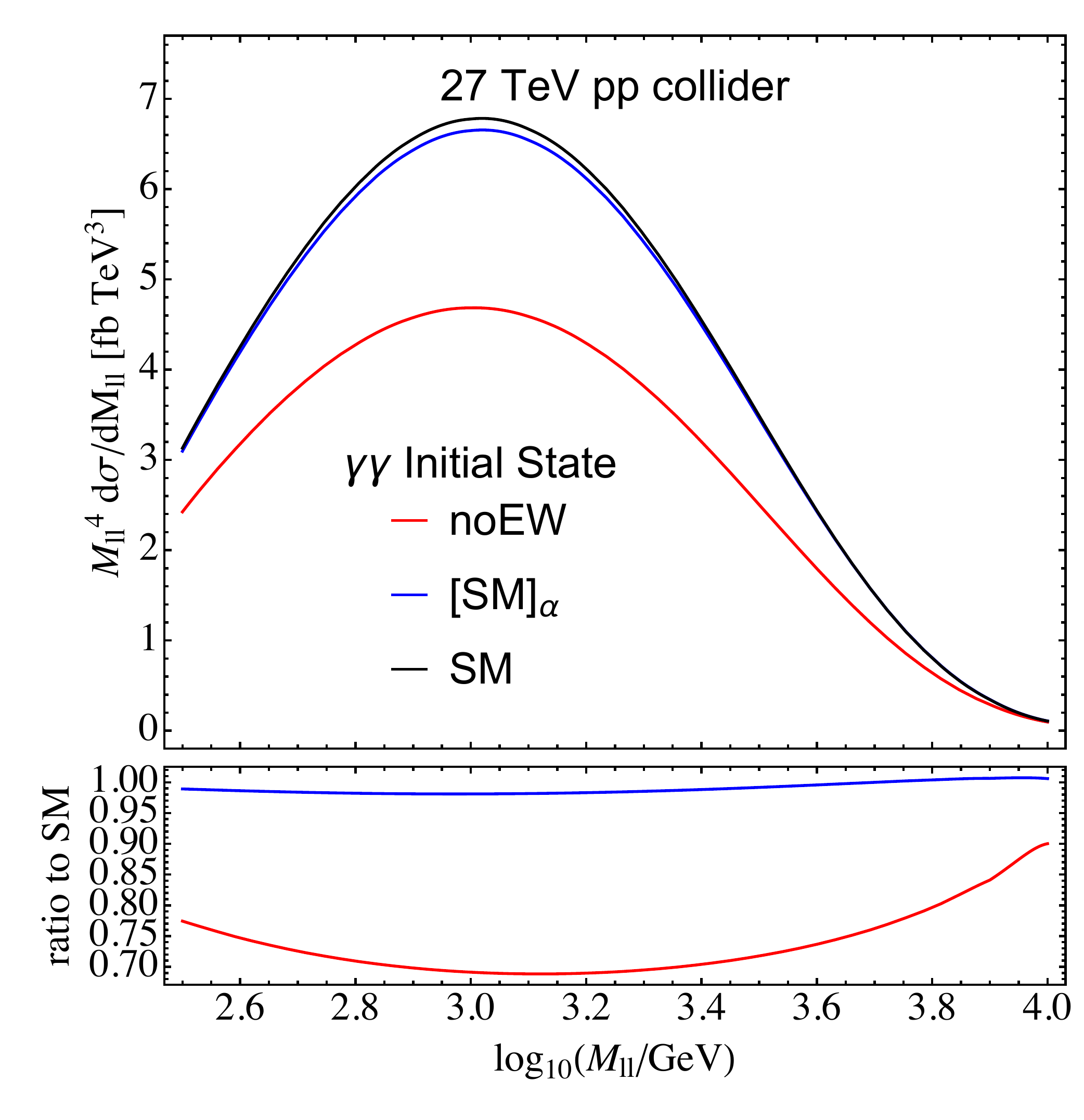}
	\includegraphics[scale=0.26]{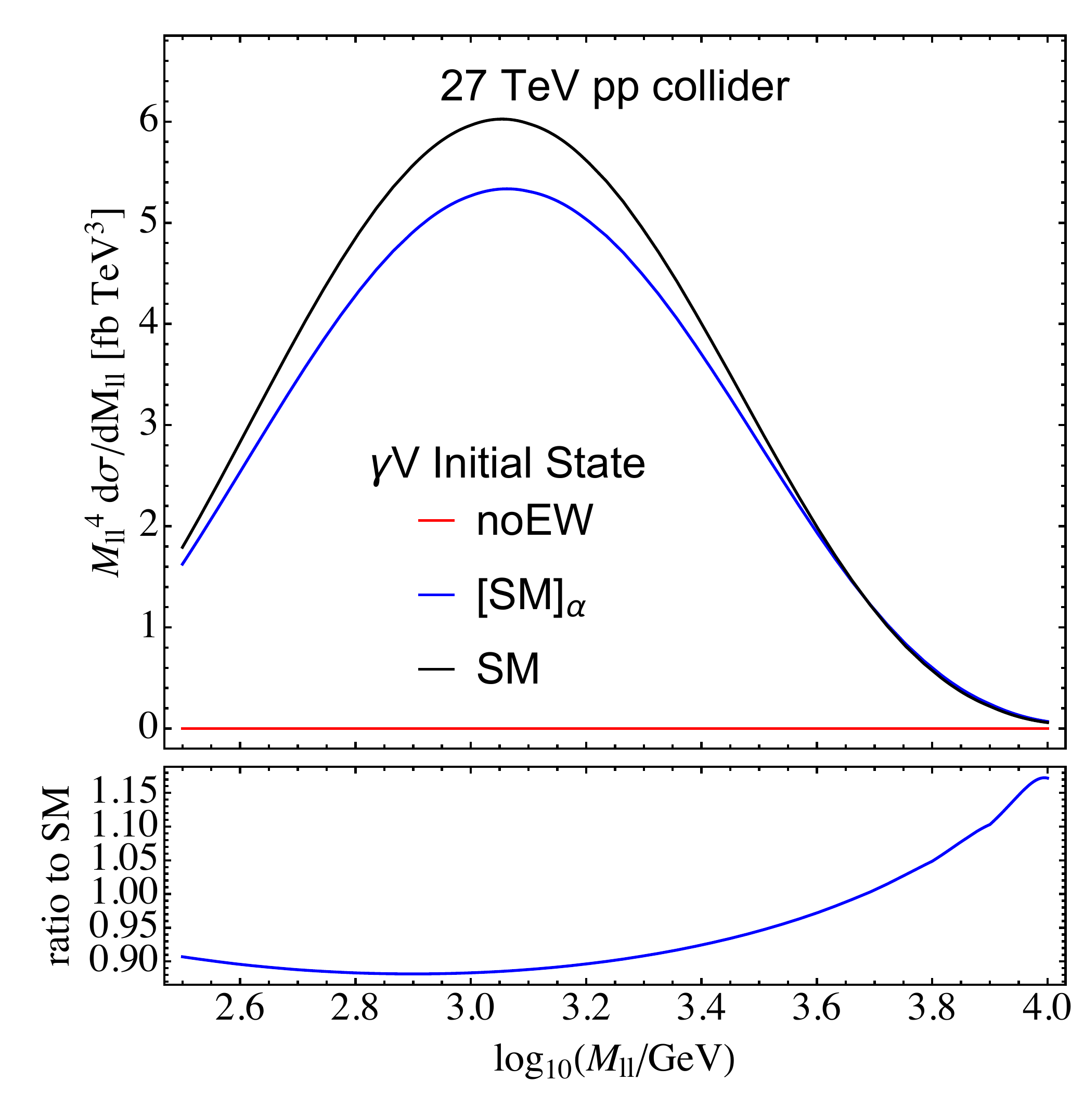}
	\includegraphics[scale=0.26]{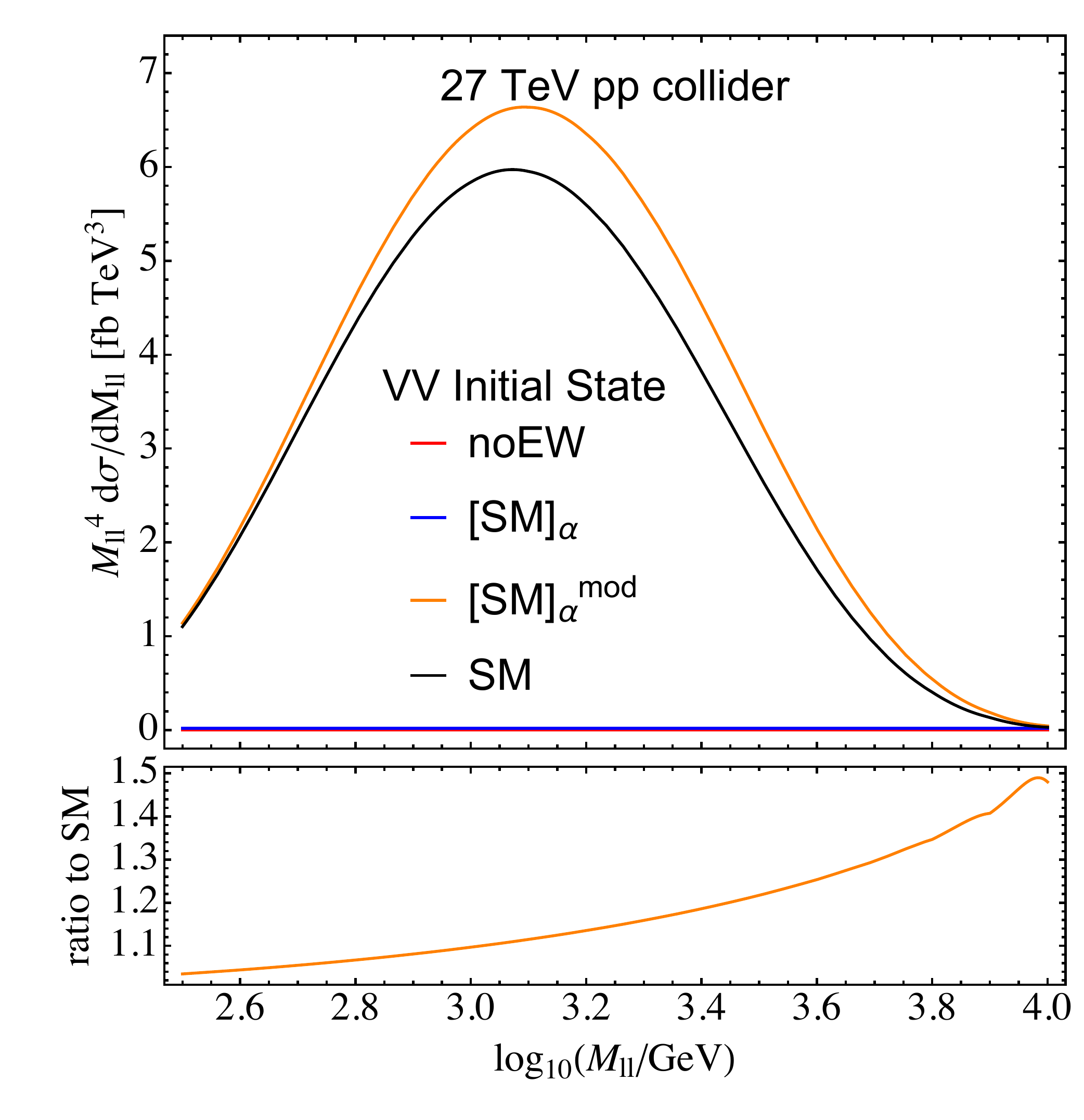}
	\caption{\label{fig:convergence_27}%
		The expansion of the various contributions to $M_{\ell\ell}^4 \df \sigma / \df M_{\ell\ell}(p_{T\ell}>100$ GeV) for a 27 TeV collider. The colors are the same as in Fig.~\ref{fig:convergence_100}.
}}
\FIGURE[h]{
 \centering
  \includegraphics[scale=0.28]{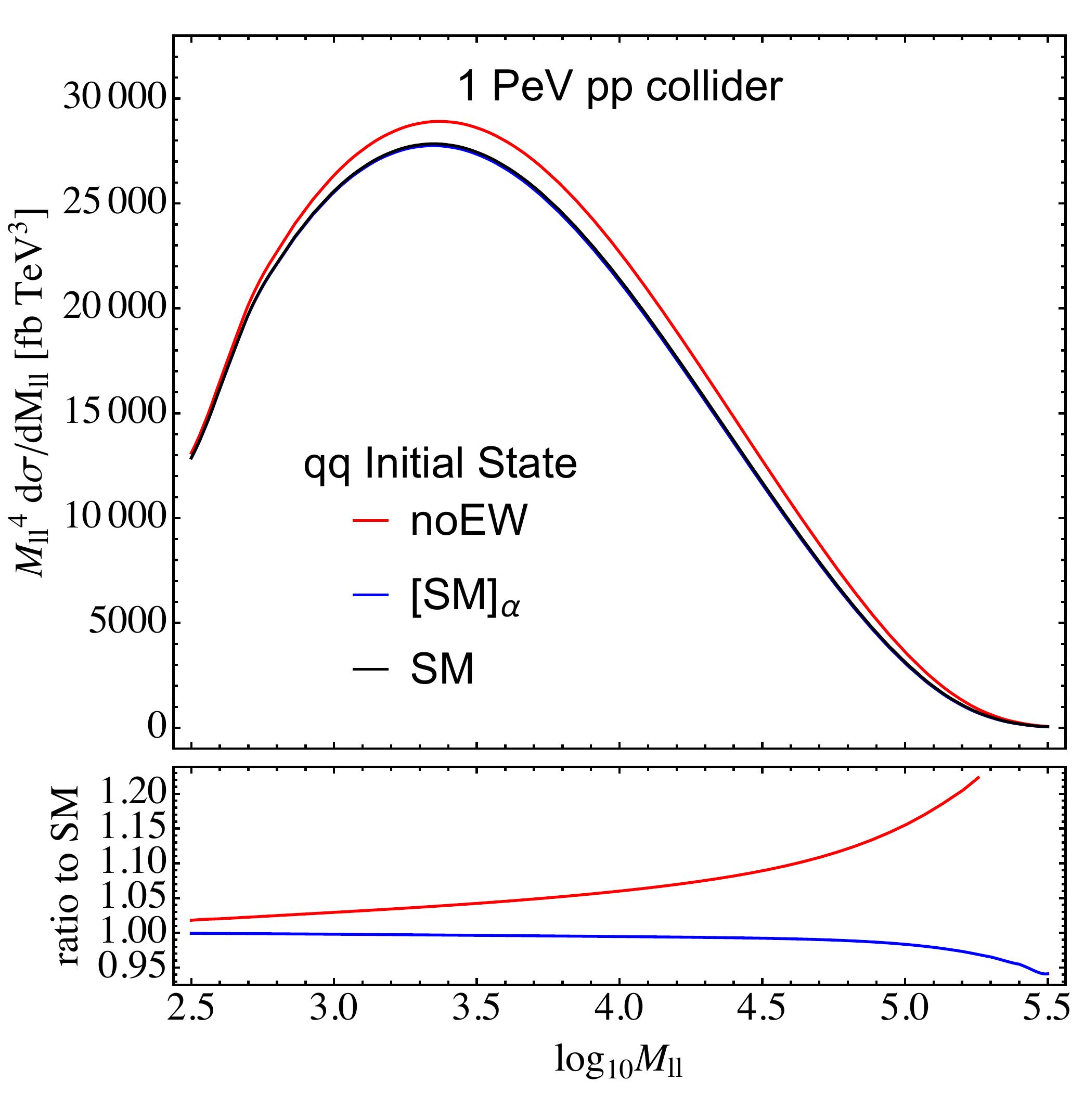}
  \includegraphics[scale=0.28]{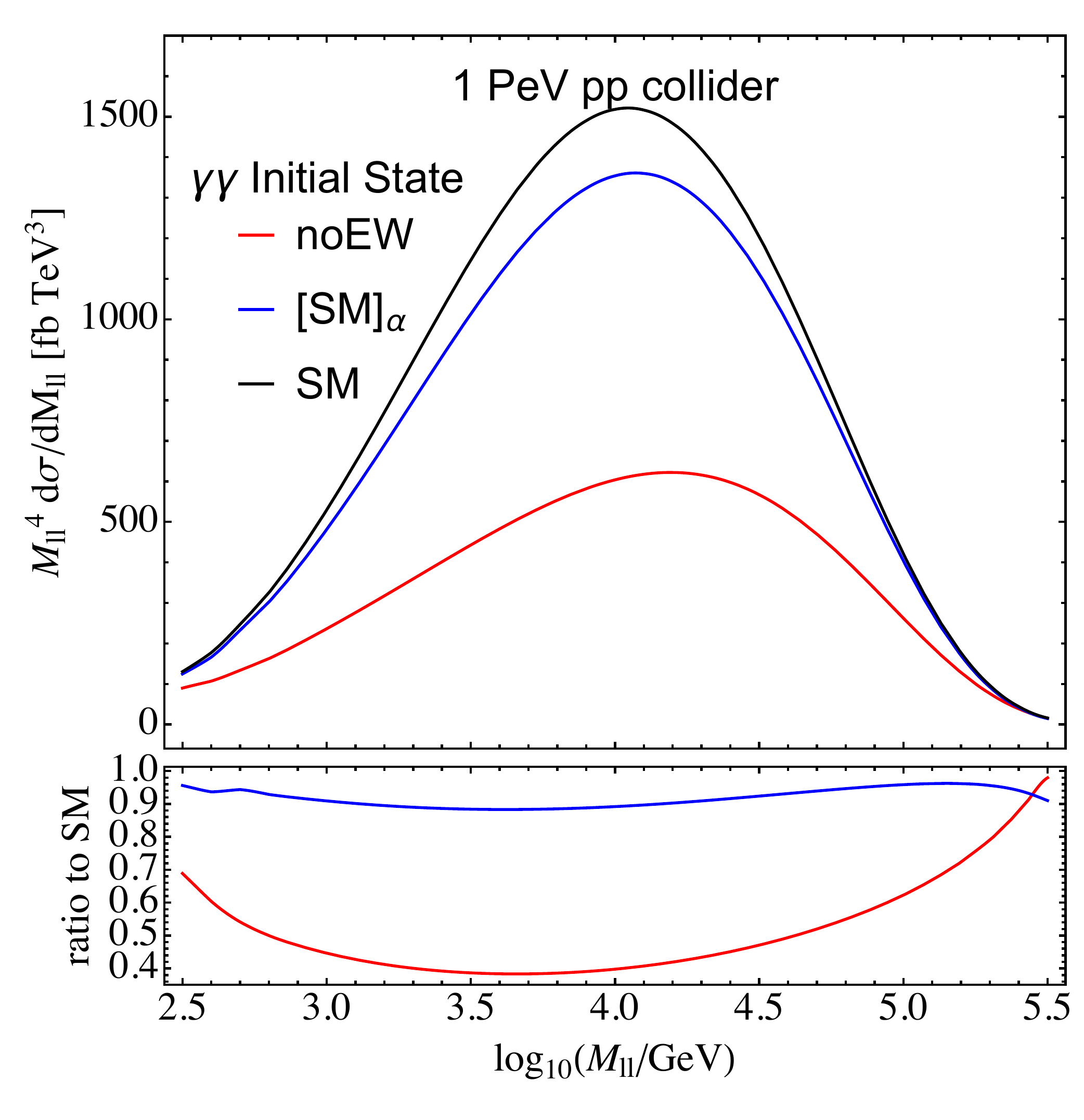}
  \includegraphics[scale=0.28]{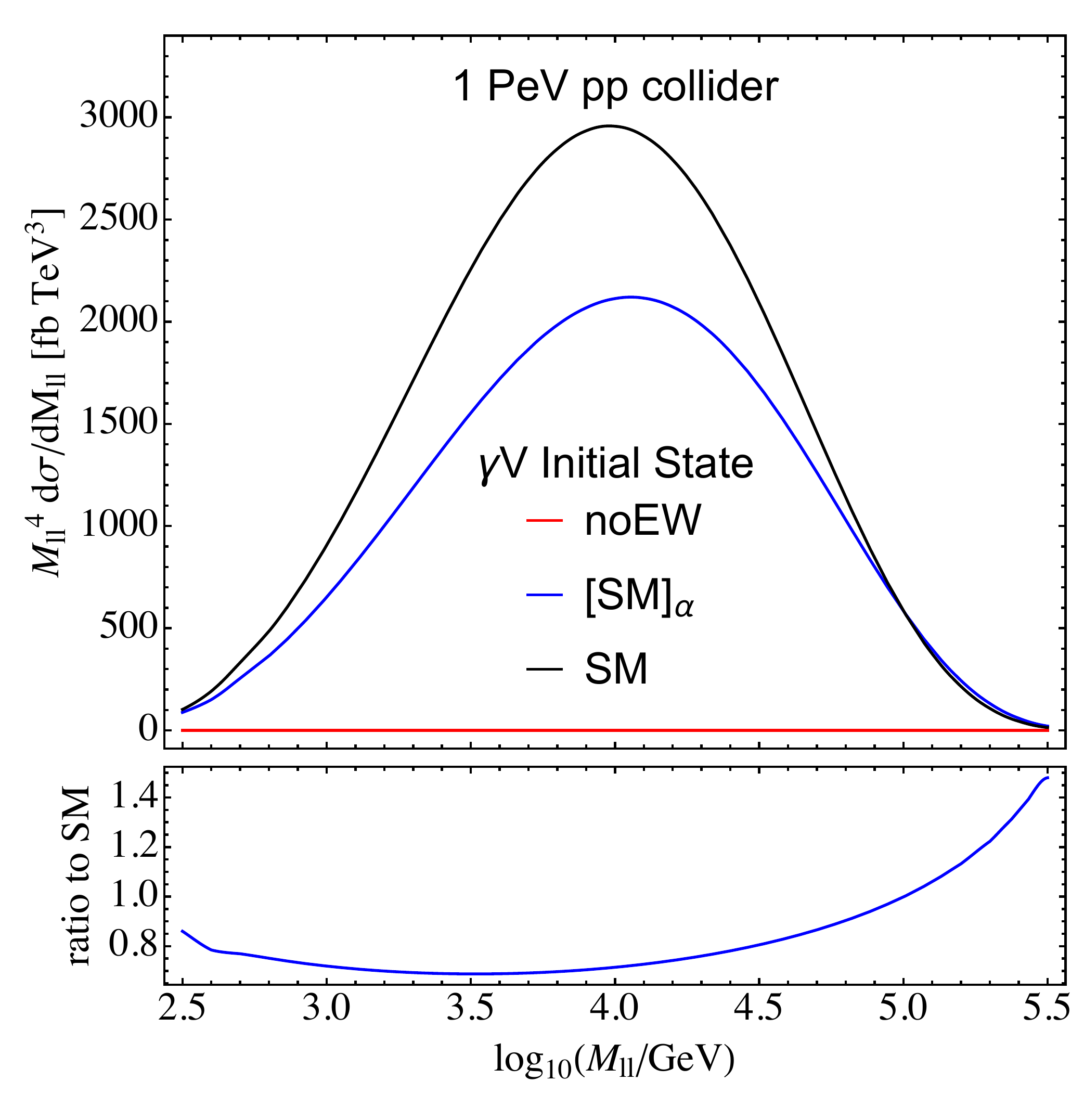}
  \includegraphics[scale=0.28]{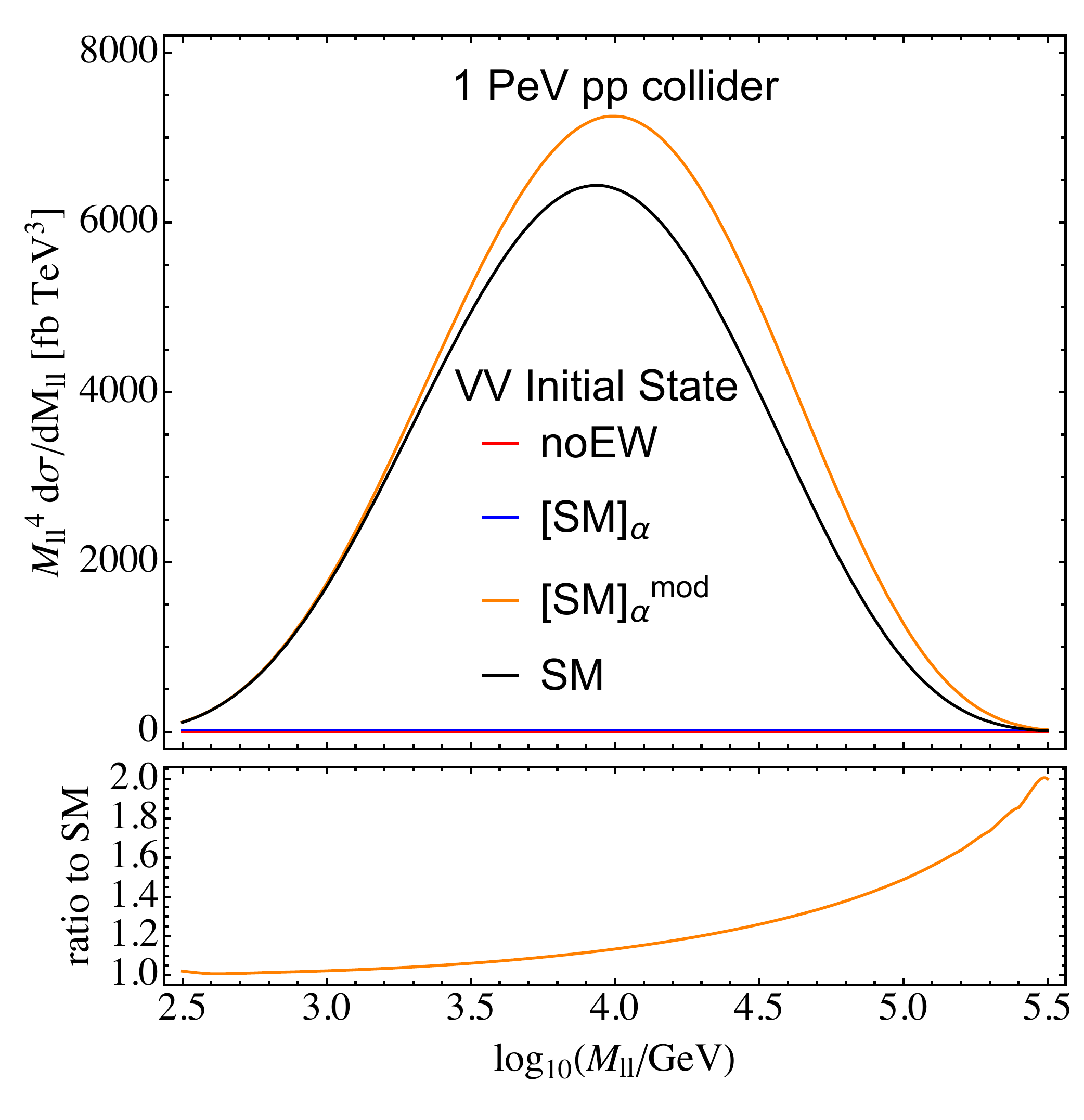}
\caption{\label{fig:convergence_1000}%
The expansion of the various contributions to $M_{\ell\ell}^4 \df \sigma / \df M_{\ell\ell}(p_{T\ell}>100$ GeV) for a 1 PeV collider. The colors are the same as in Fig.~\ref{fig:convergence_100}.
}}
\FIGURE[h]{
 \centering
  \includegraphics[scale=0.28]{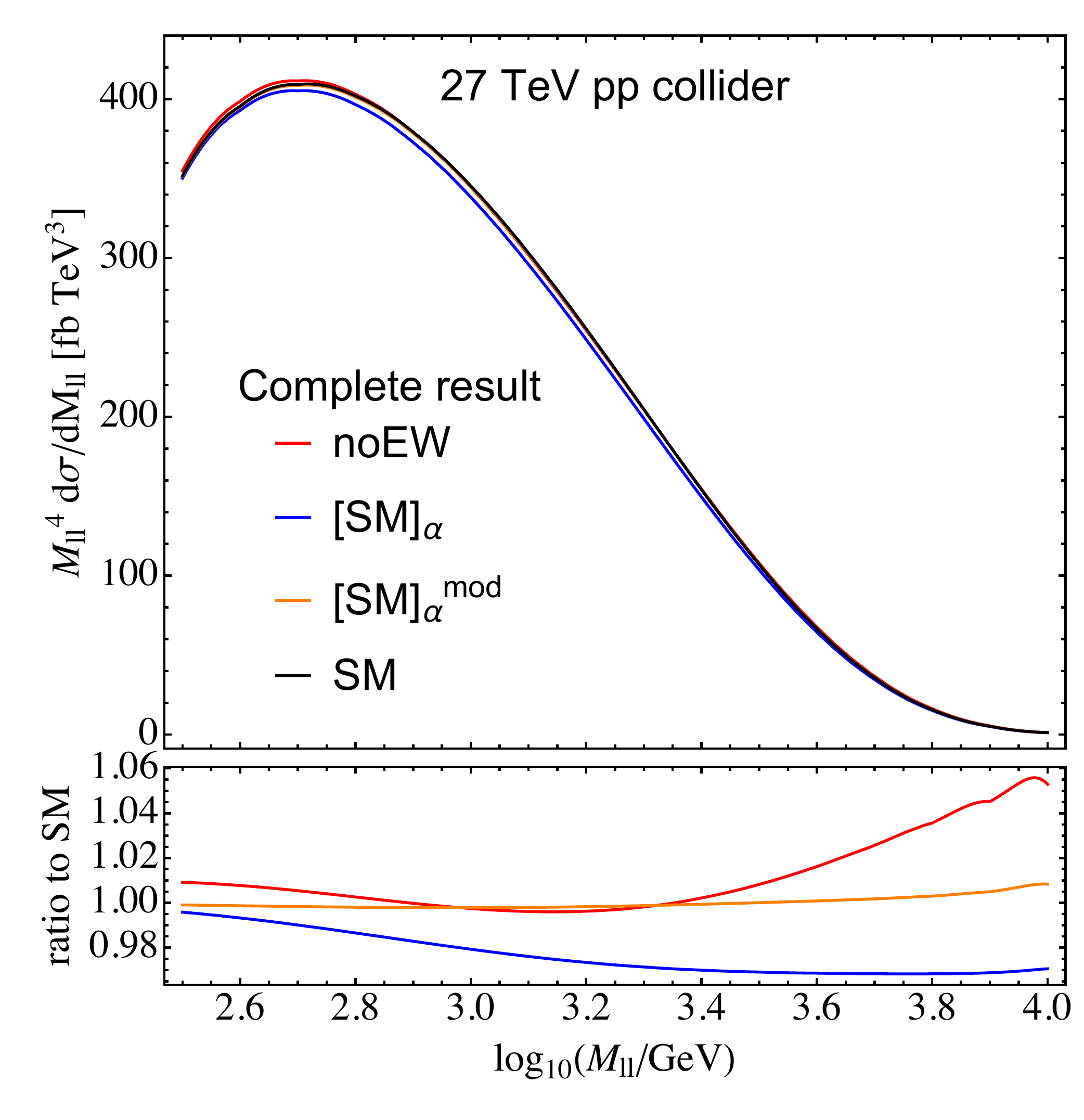}
  \includegraphics[scale=0.28]{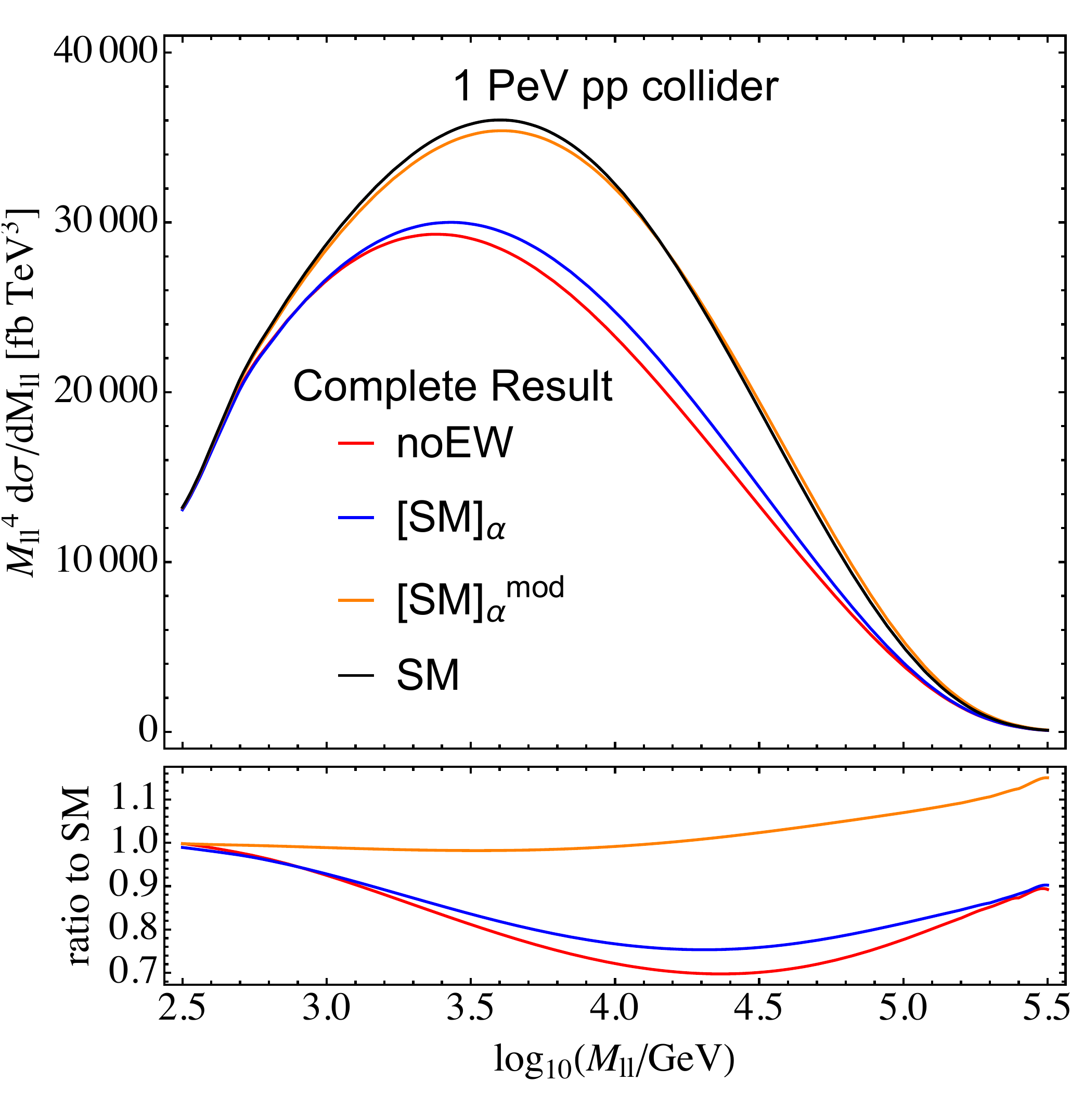}
\caption{\label{fig:total_convergence_other}%
The expansion of the complete result $M_{\ell\ell}^4 \df \sigma / \df M_{\ell\ell}(p_{T\ell}>100$ GeV) for a 27 TeV and 1 PeV collider. The colors are the same as in Fig.~\ref{fig:convergence_100}
}}

\section{Conclusions}
\label{sec:conclusions}
A fuller understanding of electroweak effects is becoming essential
as the energy frontier of particle physics moves beyond the
scale of electroweak symmetry breaking.  In this paper we have
focused on the effects of large logarithmic terms associated with
initial-state emission of electroweak bosons. Since all types of colliders
necessarily have beams that are not symmetric with respect to
weak isospin, there are double-logarithmic enhancements of
electroweak corrections associated with initial-state
radiation that do not cancel, even in fully inclusive processes, and become
increasingly important at high energies.  These enhanced terms can be
resummed to all orders by means of DGLAP-type evolution equations involving
parton distribution functions for all the fields of the Standard
Model.  The evolution equations also resum important classes of
single logarithms (but not all of them), including those associated with fermion, gluon and
U(1) gauge boson emission.

We have proposed a method for combining resummation with fixed-order
electroweak calculations, without double counting of terms already
included.  This is done by expanding the evolution equations to fixed
order in the electroweak couplings and computing the terms that need
to be subtracted to avoid double counting.  The remaining terms then
provide a resummed estimate of the higher-order effects beyond those
that have been computed exactly in fixed order.  The relative size of the first- 
and higher-order terms
provides an indication of the convergence of electroweak
perturbation theory.

In order to combine resummed and fixed-order calculations without
double counting, one needs to specify carefully the terms included in
each case.  In particular, the PDF sets used for the latter should not
include terms present in the electroweak evolution equations used for
the former.   We therefore propose a ``noEW'' scheme for fixed-order
calculations, in which there is no U(1)$_{\rm em}$ evolution above
the electroweak scale.  In particular, the photon PDF used in the
fixed-order calculation is frozen at a matching scale $q_V\sim m_V\sim
100$ GeV, and the resummation takes care of all the photon evolution above
that scale.

Using this scheme, we have presented comparisons between ``noEW''
results, the full leading-logarithmic resummation (SM) and the resummed 
results expanded to fixed order ([SM]$_\alpha$), at the level of PDFs, parton-parton
luminosities and fully-inclusive di-lepton cross sections.  The
difference between [SM]$_\alpha$ and ``noEW'' represents the part that
should be replaced by an exact order-$\alpha$ calculation for improved
precision.  The difference between SM and  [SM]$_\alpha$ then
indicates the extra contribution from the resummation of enhanced
terms of yet higher orders.  Our results are shown mainly in the
context of a future $pp$ collider of center-of-mass energy 100 TeV,
but we also show some effects at a possible 27 TeV high-energy upgrade
of the LHC and at much higher energy.

A notable feature of our findings is that there are relatively large
contributions to the PDFs of the electroweak vector bosons beyond order
$\alpha$, reaching tens of percent beyond scales of $\sim 10$ TeV.
This is reflected in their contributions to luminosities and the
di-lepton cross section. Even at fixed invariant mass, the 
relative importance of the initial states with vector
bosons increases with collider energy. This is
because at higher energies one is probing smaller values of $x$.
Since the contributions of vector boson
fusion processes begin at order $\alpha^2$, one may wish to make an
extra subtraction of this piece from the resummation, resulting in a
scheme we call  [SM]$^{\rm mod}_\alpha$.  In this scheme one can
include the exact lowest-order VBF contribution, the difference
between SM and  [SM]$^{\rm mod}_\alpha$ then indicating the effect of
remaining resummed terms.  We find that the latter are still quite
significant, again reaching tens of percent beyond scales of $\sim 10$
TeV.

Our approach naturally invites a number of future developments.
Foremost of these would be the inclusion of exact order-$\alpha$ 
calculations in the way we have proposed, together with
order-$\alpha^2$ VBF contributions.  The fully-inclusive di-lepton
process that we have considered is not experimentally relevant, owing
to the presence of unobservable neutrinos.  This could be rectified
either by including Sudakov factors for a fully exclusive
$e^+e^-$ or $\mu^+\mu^-$ final state,
or by computing fragmentation functions for the inclusive
production of charged leptons.  Ultimately, fully exclusive final
states containing all combinations of jets, leptons, photons and
massive bosons could be simulated by an event generator based
on complete Standard Model evolution equations for initial- and
final-state parton showers.

\acknowledgments
We thank Gavin Salam for helpful discussions and Tao Han,
Michelangelo Mangano and Brock Tweedie for comments on the manuscript.
This work was supported by the Director, Office of Science, Office of
High Energy Physics of the U.S. Department of Energy under the
Contract No. DE-AC02-05CH11231 (CWB, NF), and partially supported by
U.K. STFC consolidated grants ST/P000681/1 and ST/L000385/1 (BRW).

\appendix

\section{The partonic Born cross sections for di-lepton production}
\label{app:Partonic_Born}

The expressions for the Born cross sections with $AB = q \bar q$ and $AB = W^+ W^-$ are given in Table~\ref{tab:BAB_ff_charged}, 
\begin{table}
\begin{center}
\begin{tabular}{|c|c|}
\hline
$AB\to \ell\bar\ell'$ & $B_{AB}$  \\
\hline
$q_{L/R} \, \bar q_{L/R}\to \ell_{L/R} \, \bar \ell_{L/R}$ & $\frac{8\pi^2}{s} \, f_{L/R,L/R}(s,t,u) \, \left(\alpha_1 Y_q Y_\ell + \alpha_2 I_q I_\ell\right)^2$ \\
$q_{L/R} \, \bar q_{L/R}\to \ell_{R/L} \, \bar \ell_{R/L}$ & $\frac{8\pi^2}{s} \, f_{L/R,R/L}(s,t,u) \,  \alpha_1^2 Y_q^2 Y_\ell^2$ \\
$q_{L} \, \bar q_{L}'\to \ell_{L} \, \bar \ell_{L}'$ & $\frac{8\pi^2}{s} \, f_{C,L}(s,t,u) \,  \alpha_2^2$ \\
\hline
$W^+\,W^-\to e_L\bar e_L$ & $\frac{8\pi^2}{s} \,f_{(+,-)}^{(1)}(s,t,u) \,\alpha_2^2$ \\
$W^+\,W^-\to \nu_L\bar \nu_L$ & $\frac{8\pi^2}{s} \,f_{(+,-)}^{(2)}(s,t,u) \,\alpha_2^2$ \\
\hline
\end{tabular}
\end{center}
\caption{\label{tab:BAB_ff_charged} 
  Born cross sections for $q \bar q$ and $W^+ W^-$ going to lepton pairs. Here $e$ stands for the charged lepton. The cross sections for $BA \to \ell\bar\ell'$ are the same as $AB\to \ell\bar\ell'$ with $t \leftrightarrow u$.   }
\end{table}
with the various functional dependences on the Mandelstam variables $s, t, u$,
\begin{align}
s = (p_A + p_B)^2\,, \qquad t = (p_A - p_\ell)^2\,, \qquad u = (p_B - p_\ell)^2
\,,
\end{align} 
given by\footnote{In keeping with our neglect of power-suppressed
  terms above the electroweak scale, all fermion and boson masses are set to zero.} 
\begin{align}
\label{eq:fRelations1}
f_{L/R,L/R}(s,t,u) &= \frac{4}{3} \frac{u^2}{s^2}
\\
f_{L/R,R/L}(s,t,u) &= \frac{4}{3} \frac{t^2}{s^2}
\nn
f_{C,L}(s,t,u) &= \frac{1}{3}\frac{u^2}{s^2}
\nn
f_{(+,-)}^{(1)}(s,t,u) &= \frac{t}{4u} \frac{t^2+u^2}{s^2}
\nn
f_{(+,-)}^{(2)}(s,t,u) &= \frac{u}{4t} \frac{t^2+u^2}{s^2}
\nonumber\,.
\end{align}

For the scattering involving neutral gauge bosons in the initial state one can either work in the unbroken basis (where the neutral bosons required are $B$, $W_3$ or mixed $M = B/W_3$) or in the broken basis (where the neutral bosons required are $\gamma$, $Z$ or mixed $\tilde M = \gamma / Z$).
For the unbroken basis the results are given in Table~\ref{tab:BAB_unbroken} with
\begin{align}
\label{eq:fRelations2}
f_N(s,t,u) &= \frac{t^2+u^2}{ut}
\\
f_{(\pm, 3)}(s,t,u) &= \frac{1}{8} \frac{u^2+t^2}{ut} \frac{(t-u)^2}{s^2}
\nn
f_{(\pm, B)}(s,t,u) &= \frac{1}{8} \frac{u^2+t^2}{ut}
\nn
f_{(\pm, M)}(s,t,u) &= \pm\frac{1}{8} \frac{u^2+t^2}{ut} \frac{t-u}{s}
\nonumber
\,,
\end{align}
while for the broken basis the results are in Table~\ref{tab:BAB_broken} with
\begin{align}
f_{+,\gamma}(s,t,u) &=\frac 12 \frac{u^2+t^2}{s^2}\frac{u}{t}\\
f_{-,\gamma}(s,t,u) &=\frac 12 \frac{u^2+t^2}{s^2}\frac{t}{u},\nn
f_{+,Z}(s,t,u) &=\frac 18 \frac{u^2+t^2}{ut}\left(\frac{s+2c_W^2
                 u}{c_W s_W s}\right)^2\nn
f_{-,Z}(s,t,u) &=\frac 18 \frac{u^2+t^2}{ut}\left(\frac{s+2c_W^2
                 t}{c_W s_W s}\right)^2\nn
f_{+,\tilde M}(s,t,u) &=\frac 14 \frac{u^2+t^2}{st}\frac{s+2c_W^2
                 u}{c_W s_W s}\nn
f_{-,\tilde M}(s,t,u) &=\frac 14 \frac{u^2+t^2}{su}\frac{s+2c_W^2
                 t}{c_W s_W s}
                 \,,\nonumber
\end{align}
where $s_W$ and $c_W$ represent the sine and cosine of the weak mixing
angle, respectively.

\begin{table}
\begin{center}
\begin{tabular}{|c|c|}
\hline
$AB\to \ell\bar\ell'$ & $B_{AB}$  \\
\hline
$W^3\,W^3\to \ell\bar \ell$ & $\frac{8\pi^2}{s} \,f_N(s,t,u) \,\alpha_2^2 \,I_\ell^4$ \\
$W^3\,B\to \ell\bar \ell$ & $\frac{8\pi^2}{s} \,f_N(s,t,u) \,\alpha_1\, \alpha_2 \,Y_\ell^2 \,I_\ell^2$ \\
$W^3\,M\to \ell\bar \ell$ & $\frac{8\pi^2}{s} \,f_N(s,t,u)\,\sqrt{\alpha_1\alpha_2}\, \alpha_2 \,Y_\ell \,I_\ell^3$ \\
$B\,B\to \ell\bar \ell$ & $\frac{8\pi^2}{s} \,f_N(s,t,u)\,\alpha_1^2 \,Y_\ell^4$ \\
$B\,M\to \ell\bar \ell$ & $\frac{8\pi^2}{s} \,f_N(s,t,u)\,\sqrt{\alpha_1\alpha_2}\, \alpha_1 \,Y_\ell^3 \, I_\ell$ \\
$M\,M\to \ell\bar \ell$ & $\frac{8\pi^2}{s} \,f_N(s,t,u)\,\alpha_1\,\alpha_2\,Y_\ell^2 \, I_\ell^2$ \\
\hline
$W^\pm\,W^3\to \ell_L\bar \ell'_L$ & $\frac{8\pi^2}{s} \,f_{(\pm, 3)}(s,t,u) \,\alpha_2^2$ \\
$W^\pm\,B \to \ell_L\bar \ell'_L$ & $\frac{8\pi^2}{s} \,f_{(\pm, B)}(s,t,u) \,\alpha_1 \, \alpha_2$ \\
$W^\pm\,M \to \ell_L\bar \ell'_L$ & $\frac{8\pi^2}{s} \,f_{(\pm, M)}(s,t,u) \,\sqrt{\alpha_1\alpha_2} \, \alpha_2$ \\
\hline
\end{tabular}
\end{center}
\caption{\label{tab:BAB_unbroken} 
  Born cross sections for $V V$ in the unbroken basis going to lepton
  pairs. Here $M$ stands for the mixed $B/W_3$ PDF. The cross sections
for $BA \to \ell\bar\ell'$ are the same as $AB\to \ell\bar\ell'$ with $t \leftrightarrow u$.   }
\end{table}
\begin{table}
\begin{center}
\begin{tabular}{|c|c|}
\hline
$AB\to \ell\bar\ell'$ & $B_{AB}$\\ \hline
$\gamma\,\gamma\to \ell\bar \ell$ & $\frac{8\pi^2}{s} \,f_N(s,t,u) \,\alpha^2 \,Q_\ell^4$ \\
$\gamma\,Z\to \ell\bar \ell$ & $\frac{8\pi^2}{s} \,f_N(s,t,u) \,\alpha^2 \,Q_\ell^2 \,R_\ell^2$ \\
$\gamma\,\tilde M\to \ell\bar \ell$ & $\frac{8\pi^2}{s} \,f_N(s,t,u)\,\alpha^2\,Q_\ell^3 \,R_\ell$ \\
$Z\,Z\to \ell\bar \ell$ & $\frac{8\pi^2}{s} \,f_N(s,t,u)\,\alpha^2 \,R_\ell^4$ \\
$Z\,\tilde M\to \ell\bar \ell$ & $\frac{8\pi^2}{s} \,f_N(s,t,u)\,\alpha^2\,\,Q_\ell \, R_\ell^3$ \\
$\tilde M\,\tilde M\to \ell\bar \ell$ & $\frac{8\pi^2}{s} \,f_N(s,t,u)\,\alpha^2\,Q_\ell^2 \, R_\ell^2$ \\
\hline
$W^\pm\,\gamma\to \ell_L\bar \ell'_L$ & $\frac{8\pi^2}{s} \,f_{(\pm,
                                        \gamma)}(s,t,u) \,\alpha\,\alpha_2$ \\
$W^\pm\,Z \to \ell_L\bar \ell'_L$ & $\frac{8\pi^2}{s} \,f_{(\pm, Z)}(s,t,u) \,\alpha\, \alpha_2$ \\
$W^\pm\,\tilde M \to \ell_L\bar \ell'_L$ & $\frac{8\pi^2}{s}
                                           \,f_{(\pm, \tilde M)}(s,t,u) \,\alpha\, \alpha_2$ \\
\hline
\end{tabular}
\end{center}
\caption{\label{tab:BAB_broken} 
  Born cross sections for $V V$ in the broken basis going to lepton
  pairs. Here $\tilde M$ stands for the
  mixed $\gamma/Z$ PDF. The cross sections for $BA \to \ell\bar\ell'$
  are the same as $AB\to \ell\bar\ell'$ with $t \leftrightarrow u$.
}
\end{table}

\clearpage
\addcontentsline{toc}{section}{References}
\bibliographystyle{JHEP}
\bibliography{SMevol_paper}

\end{document}